\documentclass[%
 reprint,
 amsmath,amssymb,
 aps,
]{revtex4-2}

\usepackage{graphicx}% Include figure files
\usepackage{dcolumn}% Align table columns on decimal point
\usepackage[dvipsnames]{xcolor}
\usepackage{bm}% bold math
\newcommand{\zb}{\bar{z}}
\newcommand{\wb}{\bar{w}}
\newcommand{\cG}{{\cal G}}
\newcommand{\cR}{{\cal R}}
\newcommand{\cB}{{\cal B}}
\newcommand{\cX}{{\cal X}}
\newcommand{\cL}{{\cal L}}
\newcommand{\btr}{\textup{bTr}}
\newcommand{\MeijerG}[8][\bigg]{G^{{ #2 },{ #3 }}_{{ #4 },{ #5 }} #1( \begin{matrix} #6 \\ #7 \end{matrix}\, #1\vert\, #8 #1)}
\newcommand{\<}{\left<}
\renewcommand{\>}{\right>}
\def\oper{{\mathchoice{\rm 1\mskip-4mu l}{\rm 1\mskip-4mu l}
{\rm 1\mskip-4.5mu l}{\rm 1\mskip-5mu l}}}
\newcommand{\LH}{{{\rm L}}(\mathcal{H})}

\begin{document}
\title{Random generators of Markovian evolution: \\
A quantum-classical transition by superdecoherence
%Superdecoherence of random Lindblad operators
}

\author{W.~Tarnowski$^1$, I.~Yusipov$^2$, T.~Laptyeva$^2$, S.~Denisov$^{3}$,
 D.~Chru{\'s}ci{\'n}ski$^4$, and K.~\.{Z}yczkowski$^{1,5}$}
	\address{$^1$ Institute of Theoretical Physics,
Uniwersytet Jagiello\'{n}ski, 30-348 Krak{\'o}w, Poland}
	\address{$^2$ Mathematical Center, Lobachevsky University, 603950 Nizhni Novgorod, Russia}
	\address{$^3$ Department of Computer Science, Oslo Metropolitan University, N-0130 Oslo, Norway}
\address{$^4$ Institute of Physics, Faculty of Physics, Astronomy
and Informatics \\  Nicolaus Copernicus University,
%Grudzi{a}dzka 5/7,
87--100 Toru{\'n}, Poland}
\address{$^5$ Centrum Fizyki Teoretycznej PAN, 02-668 Warszawa, Poland}		
	
\vspace{10pt}

\date{\today}% It is always \today, today,

\begin{abstract}
Continuous-time Markovian evolution appears to be manifestly different in classical and quantum worlds. We consider 
ensembles of random
generators of $N$-dimensional  Markovian evolution, quantum and classical ones,  and evaluate their universal spectral properties. We then  show how the two types of generators can be  related by superdecoherence. In analogy with the  mechanism of decoherence,  which transforms a quantum state into a classical one, superdecoherence can be used to 
transform a  Lindblad operator (generator of quantum evolution) into a  Kolmogorov operator (generator of  classical evolution). We inspect spectra of random Lindblad operators undergoing superdecoherence and demonstrate that,
in the limit of complete superdecoherence, 
the resulting operators exhibit spectral density typical to random Kolmogorov operators. By gradually increasing strength of  superdecoherence, 
we observe a sharp quantum-to-classical  transition. 
Furthermore, we define an inverse procedure of  supercoherification  that is a generalization of the  scheme used to construct a quantum state out of a classical one. Finally, we study microscopic correlation between neighbouring eigenvalues through the complex spacing ratios and observe the  horse-shoe distribution, emblematic of the Ginibre universality class, for both types of random generators. Remarkably, it survives superdecoherence and supercoherification.

\end{abstract}

\maketitle

%{\sl Dedicated to memory of  Prof. Andrzej Kossakowski 

%\hskip 6.3cm (1938-2021)}

%{\color{blue} [dedication in preprint version only]}

%\section{\label{sec:level1}First-level heading:\protect\\ 
\section{Introduction}\label{sec:1}

%Quantum 
Coherence is a characteristic feature of quantum mechanics and it is now considered a valuable resource with many potential applications~\cite{coh1,coh2}. Being valuable, it is also fragile and the mechanism of decoherence, an inevitable effect of interaction of a quantum system with environment, causes loss of coherence. There is an agreement that decoherence is the key mechanism responsible for the quantum-classical transition \cite{decoh1, decoh2a, decoh2b}.

%Aside from straightforward attempts to
%go into the limit $\hbar \rightarrow \infty$ and to
% follow the coarse-graining approach \cite{man}, a %gradual increase of the decoherence 
%is considered~\cite{decoh3,decoh4,decoh5} as a means of %inducing a transition from quantum~\cite{alicki} to %classical Markovian dynamics~\cite{classic}. Classical %evolution appears then as approximation of the original %quantum one in the limit of strong %decoherence~\cite{decoh3,decoh4}.  

Quantum coherence is defined with respect to a given orthonormal basis in the Hilbert space.
Assuming that the basis is distinguished as the eigenbasis of a given Hamiltonian, the quantum coherences are encoded in the off-diagonal elements of a
 density operator $\rho_{ij} = \<i |\rho|j\>$. 
 The simplest decoherence process is  realized by the action of the coarse-graining  quantum channel,
\begin{equation}\label{11}
  \rho \to \Delta^{(p)}(\rho) = \sum_{i,j} \Delta_{ij} \rho_{ij} |i \rangle \langle j| ,
\end{equation}
such that $\Delta_{ii}=1$, and the role of off-diagonal factors $\Delta_{ij} = p < 1$ is to suppress off-diagonal elements $\rho_{ij}$.

Assume now that a certain basis $|i\rangle$ in the Hilbert space $\mathcal{H}$  is distinguished by an interaction of a system with an environment.  In the case of quantum maps (or channels, if the maps are trace-preserving)~ \cite{res1,res2,Watrous},   {\em superdecoherence} can be defined as decoherence acting on the states related to quantum maps by the Jamio\l{}kowski-Choi isomorphism \cite{jam,choi},  also known as the `channel-state duality'~\cite{duala,duality}.

Any quantum map  $\Phi$ acting on  a system of size $N$  can be represented  by a Choi matrix  $\mathbf{C}$
 of order $N^2$,
\begin{equation}\label{dc}
 % \mathcal{E}(\rho) = \sum_{i,j} \sum_{k,l} \mathcal{E}_{ij,kl} |i  \rangle \langle j| \, \rho \, |l \rangle \langle k| .  
 %% change of notation to  be consistent with section 2
\rho \to 
% \rho'=
\Phi(\rho) = \sum_{i,j} \sum_{k,l}
 \mathbf{C}_{ij,kl} \, |i  \rangle \langle j| \, \rho \, |l \rangle \langle k| .
\end{equation}

The process of superdecoherence is similarly realized by the action of a supermap
$\widetilde{\Delta}$,  which acts on maps \cite{CDAP08,Zy08}
and  suppresses the off-diagonal elements of   %$  \mathcal{E}_{ij,kl}$,
$\mathbf{C}_{ij,kl}$, 
\begin{eqnarray}
\label{superdeco}
\Phi (\rho) \to \widetilde{\Delta}^{(p)}\bigl(\Phi(\rho) \bigr) = ~~~~~~~~~~~~~~~~~~~~~~~~~ \nonumber \\ 
= \sum_{i,j} \sum_{k,l} \widetilde{\Delta}_{ij,kl}  
\mathbf{C}_{ij,kl} \,
|i  \rangle \langle j| \, \rho \, |l \rangle \langle k|, ~~~~
\end{eqnarray}
where diagonal elements remain unchanged, $ \widetilde{\Delta}_{ij,ij} = 1$, 
while the off-diagonal entries become  suppressed,  $ \widetilde{\Delta}_{ij,kl} = p \leq 1$. Clearly, in the limit $p \to 0$ one arrives at a map which completely destroys coherence of any input state $\rho$
\begin{equation}\label{}
 %\mathcal{E} \to \widetilde{\Delta}^{(0)}(\mathcal{E})(\rho) = \sum_{i,j} S_{ij} |i  \rangle \langle j| \, \rho \, |j  \rangle \langle i| ,
 \Phi(\rho) \to \widetilde{\Delta}^{(0)}
 \bigl(\Phi(\rho) \bigr) 
   = \sum_{i,j} S_{ij} \, |i  \rangle \langle j| \, \rho \, |j  \rangle \langle i| ,
\end{equation}
where $S_{ij} = \mathbf{C}_{ij,ij}$ denotes
a stochastic matrix representing the completely decohered quantum channel $\Phi$.
By continuing further along this line, as we show in this work, one can define superdecoherence for generators of continuous-time Markovian quantum evolution of the Gorini-Kossakowski-Sudarshan-Lindblad  (GKSL) form \cite{gorini,lindblad,CP17} (often called  Lindblad operators or simply \textit{Lindbladians} \cite{alicki}), 
which decohere them into classical Kolmogorov operators, governing classical evolution inside the probability simplex.

Instead of analyzing a particular, well-defined physical system, 
one can ask about properties of a \textit{typical} system. It is then advantageous
to rely on the concepts of random matrix theory (RMT) \cite{Mehta}, which is capable of describing the behavior of a typical quantum system 
under the assumption that its dynamics is related to a
classically chaotic system.
%which displays generic  behavior. 
Such an approach emerged as a natural way 
to deal with many-body problems in nuclear physics \cite{Wigner} 
and it was well established by a series of works on statistical theory of spectra by Dyson~  \cite{Dyson1,Dyson2,Dyson3,Dyson4}. 

Due to a seminal monograph by Haake \cite{Haake1} 
and his collaborators \cite{Haake4}, random matrices became a key theoretical tool to study links between classical and quantum chaotic systems (see  also Refs.~\cite{BRAUN,CHAOS}). 
Random matrices found interesting applications in the analysis of certain models of 2D quantum gravity \cite{a1,a2}, gauge theories \cite{a3},
 the theory of open quantum systems \cite{open-random}.
Moreover, they played an essential role in tackling key problems in quantum information theory \cite{qi2,qi1}.

In our recent work  \cite{denysov2019} we introduced an ensemble of random Lindblad operators.
%, and looked for
%characteristics of a generic quantum evolution in %continuous time.
We found  universal spectral properties of typical, i.e., randomly sampled, Lindblad operators. It was shown that, in the case of purely dissipative dynamics, their spectral distributions have a universal lemon--like shape on the complex plane. 

Spectral properties of random Lindblad generators with various types of randomness were also analyzed in recent works~\cite{Ca19,COOG19,SRP20,Prosen-PRX2020,WPL20,LT21}.
The interest to this problem was sparked by the idea to generalize the existing theory of Quantum Chaos -- which is based on spectral properties of generators of unitary evolution, i.e., Hamiltonians \cite{Haake4} - to the case of open many-body systems (see also recent works~\cite{campo1,campo2}). Spectral properties of random purely dissipative Lindbladians, on which some  additional constrains were imposed, were studied in a recent work~\cite{SPL2020},  by implementing random, identity-equivalent (in the unitary limit) circuits on the IBM Quantum platform~\cite{IBM}.

A similar problem can also be considered for generators  of classical Markovian evolution \cite{S13,classic}. Spectra of random Markov transition matrices were studied in Ref.~~\cite{S14}. Spectra of random Kolmogorov operators were first analyzed in Ref.~\cite{timm} and later in Ref.~\cite{LT21}.  Kolmogorov generators sampled from an ensemble of random graphs and their spectral properties were studied in Ref.~\cite{LT21}.

The aim of this work 
%dedicated to spectral properties of %generic Lindblad operators
is two-fold. On the one hand, we extend the previous results which addressed 
 the support of the spectrum~\cite{denysov2019} only, and now focus on the universal \emph{spectral density} $\rho(z,\zb)$.

On the other hand,  using the tools of random matrix theory ~\cite{Mehta},
we analyze the quantum--to--classical transition  at the level of the spectrum of typical Lindblad operators,  which are decohered into Kolmogorov operators.  This transition is realized with the  supermap $\widetilde{\Delta}^{(p)}$,   Eq.~(\ref{superdeco}).
We demonstrate that in the limit  $p \to 0$  one perfectly recovers an ensemble of random Kolmogorov operators. However, to derive a universal pattern for the transitions of the support of the spectrum of typical generators acting on systems of an arbitrary size $N$, 
one has to use  a different scaling in the quantum (near $p=1$) and in the classical (near $p=0$) regimes. Namely, in the quantum case spectra have to be scaled with $N/p$, whereas in
the classical case with $N^{3/2}$.  
We evaluate scaling characteristics of  the  quantum to classical  cross-over  and demonstrate a kind of a phase transition which occurs at   the critical value  $p_{\mathrm{tr}} = N^{-1/2}$,
similar to a recently reported in the case of superdecoherence of  quantum operations  %\cite{qi3}. 
\cite{KNPPZ21}.

This paper is organized as follows. In Section~\ref{Preliminaries}
we introduce the necessary concepts and notations.
Spectra of random Lindbladians are studied in Section~\ref{sec:2},
in which the spectral density is also evaluated.
In Section~\ref{sec:2c} we analyze 
spectra of random generators of classical Markovian evolution,
while quantum-to-classical transition on the level of generators is investigated in Section~\ref{sec:3}. 
In Section~\ref{sec:4} we introduce and discuss the procedure of coherification that can be used to transform a Kolmogorov operator into a Lindblad operator.  Finally we make
concluding remarks in Section~\ref{sec:5}. We present technical details in the Appendixes, including a detailed derivation of spectral densities, both for random quantum and classical generators, obtained with methods of  free probability.

\section{Setting the scene}
\label{Preliminaries}

\begin{figure*}[t]
\begin{center}
\includegraphics[angle=0,width=0.99\textwidth]{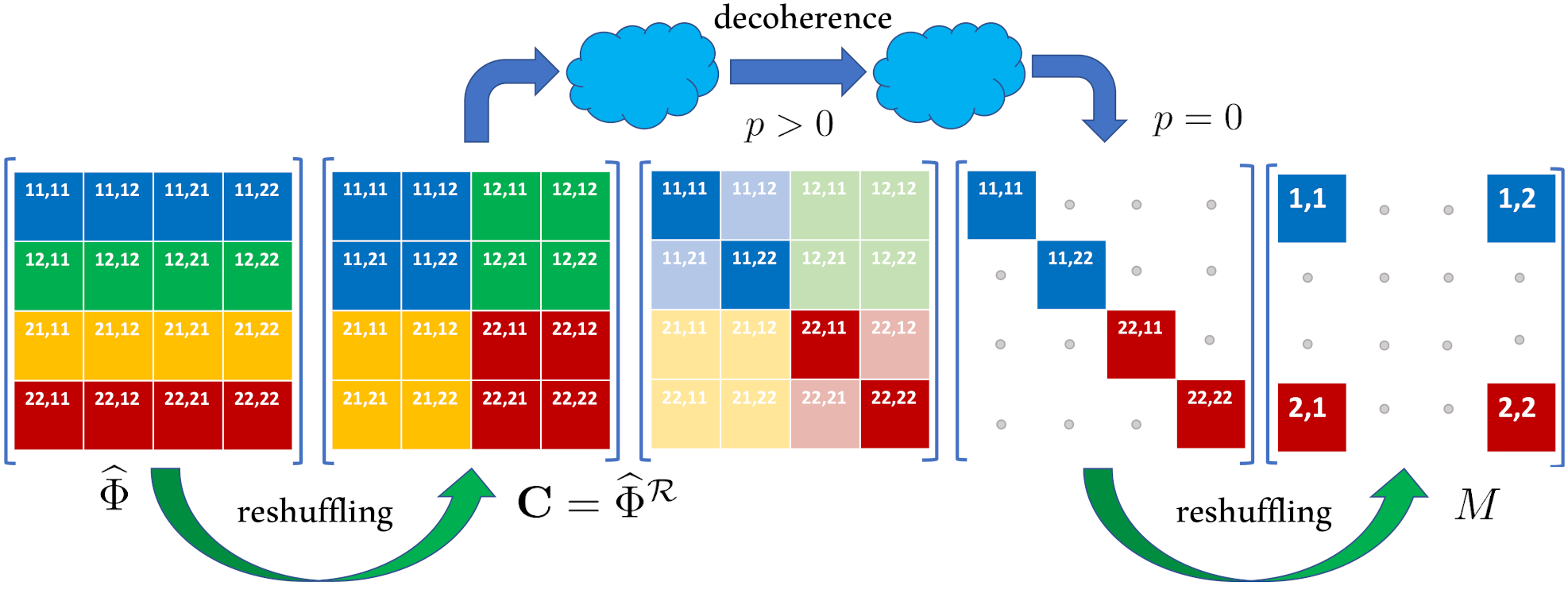}
%\end{center}	
\caption{Superdecoherence acting on a completely positive map for $N=2$. The map is specified with superopeator matrix $\widehat{\Phi}$, Eqs.~(\ref{super1}-\ref{super2}).
Next, reshuffling operation, Eq.~(\ref{resh}), is  performed on $\widehat{\Phi}$ and Choi matrix $\mathbf{C}$ is obtained. A tunable decoherence, Eq.~(\ref{dc}), then acts on the Choi matrix and diminishes its off-diagonal elements. 
In the limit of complete decoherence, $p=0$, only the diagonal elements survive, so
after the second  reshuffling
and projecting out two central rows and columns,
which describe the evolution of coherences,
one arrives at the matrix $M$
of order $2$ with non-negative entries.
%Note that the original map does not need to be trace-preserving and does not have a specific structure which makes $M$ a transition matrix.
If the original quantum map 
$\widehat{\Phi}$
is stochastic (bistochastic),
then the transition
matrix $M$ is 
 stochastic (bistochastic).}
\end{center}	
\label{fig:Fig0}
\end{figure*}

Consider a linear map $\Phi: \LH \to \LH$, where $\LH$ denotes the vector space of linear operators acting on the Hilbert space $\mathcal{H}$. We assume that ${\rm dim}\, \mathcal{H} = N$.  In this work we consider only  Hermiticity preserving maps, i.e., $\Phi(X)^\dagger = \Phi(X^\dagger)$ for all $X \in \LH$. There are several ways to find matrix representation of  $\Phi$. Fixing an orthonormal basis $\{|1\rangle, \ldots,|N\rangle\}$ in $\mathcal{H}$, one can define Choi operator %$\mathbf{C}^\Phi$ 
$\mathbf{C}$~\cite{jam,choi}
\begin{equation}\label{Choi1}
  \mathbf{C} = \sum_{i,j=1}^N |i\rangle \langle j| \otimes \Phi(|i\rangle \langle j|) ,
\end{equation}
and the corresponding $N^2 \times 
 N^2$ Hermitian  matrix
\begin{equation}\label{}
   \mathbf{C}_{ij,kl} := \langle i \otimes j|  \mathbf{C} | k \otimes l\rangle  = \langle j| \Phi( |i\rangle \langle k|)| l\rangle .
\end{equation}
The map $\Phi$ is completely positive if and only if $ \mathbf{C} \geq 0$. Another useful representation is defined as follows: Any matrix  $X \in \LH$ may be mapped to a vector  $|X \rangle \! \rangle \in \mathcal{H} \otimes \mathcal{H}$ as follows
\begin{equation}\label{}
  |X \rangle \! \rangle = \sum_{i,j=1}^N X_{ij} |i \otimes j\rangle ,
\end{equation}
where $X_{ij} = \langle i|X|j\rangle$. Operation $X \to |X \rangle \! \rangle$ is often called `vectorization' \cite{Watrous,Gilchrist}. It allows one to define a super-operator  $\widehat{\Phi} \in L(\mathcal{H} \otimes \mathcal{H})$ via
\begin{equation}\label{super1}
  \widehat{\Phi}  |X \rangle \! \rangle :=    |\Phi(X) \rangle \! \rangle ,
\end{equation}
and the corresponding matrix reads
\begin{equation}\label{super2}
   \widehat{\Phi}_{ij,kl} := \langle i \otimes j|  \widehat{\Phi} | k \otimes l\rangle .
\end{equation}
These two matrix representations are related by the reshuffling operation \cite{res2}
\begin{equation}\label{resh}
\mathbf{C} = \widehat{\Phi}^\mathcal{R},~~~ \mathbf{C}_{ij,kl}  = 
   \widehat{\Phi}_{ik,jl} .
\end{equation}
Note that $ \widehat{\Phi}_{ij,kl}$, contrary to 
$\mathbf{C}_{ij,kl}$,  is not Hermitian. If $\Phi$ is completely positive and
\begin{equation}\label{}
  \Phi(X) = \sum_\alpha K_\alpha X K_\alpha^\dagger
\end{equation}
denotes its Kraus representation, then
\begin{equation}\label{}
   \mathbf{C} = \sum_\alpha |K_\alpha   \rangle \! \rangle  \langle \! \langle K_\alpha|
\end{equation}
and
\begin{equation}\label{}
 \widehat{\Phi} =  \sum_\alpha K_\alpha \otimes \overline{K}_\alpha.
\end{equation}
Interestingly, the spectrum of the map $\Phi$ coincides with a spectrum of the super-operator $\widehat{\Phi}$, that is, $\Phi(X) = \lambda X$ if and only if $\widehat{\Phi} |X \rangle \! \rangle = \lambda |X\rangle \! \rangle$.

A third useful representation is provided by the following real $N^2 \times N^2$ matrix: let $\tau_\alpha$ $(\alpha=0,1,\ldots,N^2-1)$ denote a Hermitian orthonormal basis in $\LH$, that is, ${\rm Tr}(\tau_\alpha \tau_\beta) = \delta_{\alpha\beta}$, and let $\tau_0 = \oper/\sqrt{N}$. Define
\begin{equation}\label{TILDE}
   \widetilde{\Phi}_{\alpha\beta} = {\rm Tr}(\tau_\alpha \Phi(\tau_\beta)) ,
\end{equation}
which is by construction a real matrix. Note, that if $\Phi$ is trace-preserving, then $ \widetilde{\Phi}_{\alpha\beta}$ has the following structure
\begin{equation}\label{fano}
   \widetilde{\Phi}_{\alpha\beta} = \begin{pmatrix}
                                      1 & 0 \\
                                      \mathbf{x} & \mathrm{T}
                                    \end{pmatrix} ,
\end{equation}
where $\mathbf{x} \in \mathbb{R}^{N^2-1}$, and $\mathrm{T}$ is
a real square matrix of order 
 $N^2-1$. In this case the spectrum of $\Phi$ consists of the leading eigenvalue equal to $1$ and the spectrum of $T$. It is well known that the spectra of $\Phi$, $\widehat{\Phi}$ and $\widetilde{\Phi}$ coincide. On the one hand, the form (\ref{fano}) is called the  'Liouville  representation'  \cite{KSRJO14} of  map $\Phi$. It can 
 be also related with the
Fano form \cite{Fa83} of the bi-partite state representing the map through the
Choi-Jamio{\l}kowski isomorphism.
%}

Now, when we related every completely positive map $\Phi$ to 
unique Choi state $\mathbf{C}$, the superdecoherence acting on the maps can be defined in the straightforward manner summarized on Figure~1.

Various  ensembles of quantum states analyzed in Ref.~\cite{ZPNC11} proved to be useful for constructing ensembles of random channels \cite{KNPPZ21}.
It was shown that spectral properties of superoperators
corresponding to random quantum channels \cite{BCSZ09},
acting on a system of size $N$,
describe universal features of spectra of interacting quantum chaotic systems.
The spectrum of a typical random superoperator
consists of the leading Frobenius-Perron eigenvalue, $\lambda_1=1$, 
and the bulk of complex eigenvalues located on the origin-centered disk of radius $R\sim 1/N$ \cite{BSCSZ10}.
The spectrum of generic stochastic matrices
has a similar structure \cite{S14}, but the radius of the eigenvalue disk  depends on the measure used for  generating the ensemble~\cite{KNPPZ21}.

\section{Spectra of random Lindblad operators}
\label{sec:2}

In this section we discuss universal spectral features of typical (random) Lindblad operators.

\subsection{Ensembles of random Lindbladians}\label{sec:2q}

\begin{figure*}[t]
\begin{center}
	\includegraphics[angle=0,width=0.99\textwidth]{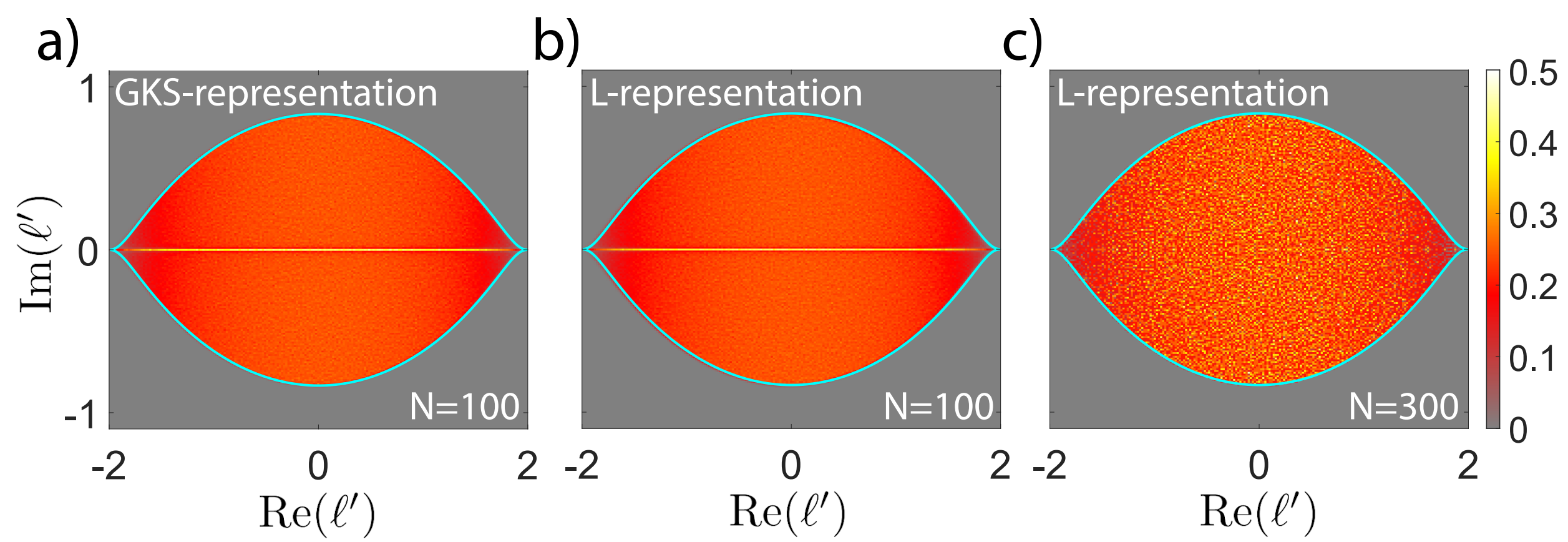}
%\end{center}	
\caption{Probability density functions
$P[\mathrm{Re}(\ell'),\mathrm{Im}(\ell')]$
of the rescaled eigenvalues,	$\ell'= N (\ell +1)$,
of random Lindblad operator  ${\cal L}$ for $N=100$ and $N=300$ (a single sample), sampled according to the (a) Gorini-Kossakowski-Sudarshan, Eq.~(\ref{L0}), and (b-c) Lindblad,
Eq.~(\ref{L1}), representations.  Densities for $N=100$ were sampled with $10^3$  realizations. Bright contour corresponds to the spectral boundary, Eqs.~(\ref{eq:BorderlineFinalM}-\ref{eq:contour2}). Note that the eigenvalue $\ell_1 = 0$ is excluded from the plots.}
\end{center}	
	\label{fig:1}
\end{figure*}

Consider a generator $\mathcal{L}$ of quantum dynamical semigroup \cite{alicki}. It has the well-known  Gorini-Kossakowski-Sudarshan-Lindblad form \cite{gorini,lindblad}
\begin{equation}\label{L}
  \mathcal{L}(\rho) = -i[H,\rho] + \sum_k \gamma_k \Big(L_k \rho L_k^\dagger - \frac 12 \{ L_k^\dagger L_k,\rho\} \Big),
\end{equation}
where $H^\dagger = H$ stands for the system Hamiltonian, $L_k$ are Lindblad (jump) operators, and all rates are positive, $\gamma_k > 0$. In what follows, we keep $\hbar = 1$. The above representation is  not unique. In particular, the splitting into the Hamiltonian part and dissipative parts can be performed in many equivalent ways. Fixing orthonormal basis $F_k$ in $\LH$ such that $F_0 = \oper/\sqrt{N}$, one finds  the following  representation \cite{gorini}:
\begin{eqnarray}  \label{L0}
\mathcal{L}(\rho) = -i[H_0,\rho] + ~~~~~~~~~~~~~~~~~~~~~~~~~~~~~~~~~~~~~~~\nonumber \\ \sum\limits_{m,n=1}^{N^2-1} \!\! K_{mn} \Big[F_n\rho F^{\dagger}_m - \frac 12 (F^{\dagger}_m F_n\rho+\rho F^{\dagger}_m F_n)\Big].~~~
\end{eqnarray}
Note that $F_k$ are in general non-Hermitian and ${\rm Tr} F_k =0$ for $k=1,\ldots,N^2-1$. Moreover, by requiring that $H_0$ is traceless, the representation (\ref{L0}) is made unique.   
The {\em Kossakowski matrix}  $K = \{K_{mn}\}$, which  specifies the dissipative part of 
$\mathcal{L}$, is positive semi-definite. Since the form (\ref{L0}) was proposed by Gorini, Kossakowski, and Sudarshan in their seminal paper \cite{gorini}, henceforth we refer to it as {\em GKS-representation}.

In recent paper \cite{denysov2019} we analyzed spectral properties of random Lindbladians. For the purely dissipative case, i.e., $H_0=0$, Lindbladian $\mathcal{L}$ is fully determined by Kossakowski $K_{mn}$ matrix. The latter can be sampled in many ways~\cite{denysov2019}.
%(see Supplementary Material in Ref.~\cite{denysov2019}).
A particular choice is not so important (provided that it is not pathological) because spectral features of the corresponding  Lindbladian ensembles are universal and for $N \gtrsim 100$
typicality emerges, i.e., a single sample yields a spectrum reproducing the universal distribution.

The most intuitive idea is to sample $K$ from the ensemble of complex Wishart matrices \cite{ZPNC11}
\begin{equation}
K = N GG^\dagger/\mathrm{Tr}GG^\dagger,
\label{eq:RandomKoss}
\end{equation}
where $G$ is a complex square Ginibre matrix with independent identically distributed (i.i.d.) complex Gaussian entries. Due to the unitary invariance of $K$, a particular choice of basis 
$\{F_n \}$ is irrelevant. For instance, in Ref.~\cite{denysov2019} we used generators of the $SU(N)$ group~\cite{alicki}.

As it was demonstrated~\cite{denysov2019}, after the scaling transformation ${\cal L}'= N ({\cal L} + \mathcal{I})$, where $\mathcal{I}$ is the identity superoperator, this sampling
results in a universal, asymptotically $N$-independent spectral distribution (probability density function) with a characteristic lemon-like shape of its support; see Fig.~2(a). We present a detailed RMT-based evaluation of the universal  distribution in the next section.

From the computational point of view, a realization of the sampling based on the representation (\ref{L0}) is a resource demanding procedure~\cite{PRE2019}. By using a parallelization technique and implementing an optimized algorithm on a cluster, it is possible to obtain samples for $N=200$ \cite{entropy2020}. This limit is determined by the complexity of the `wrapping' of the Kossakowski matrix $K$ into the elements of a  Hilbert-Schmidt basis $\{F_n\}$.
%, see~Eq.~(\ref{L0}).

Equation (\ref{L}) can be equivalently rewritten in the following compact form:
\begin{equation}\label{L1}
  \mathcal{L}(\rho) = -i[H,\rho]+ \Phi(\rho) -
  \frac 12 \bigl( \Phi^\ddag(\oper) \rho + \rho \Phi^\ddag(\oper) \bigr),
\end{equation}
where $\Phi$ is a completely positive (CP) map, $\Phi(\rho) = \sum_k \gamma_k L_k \rho L_k^\dagger$, and $\Phi^\ddag$ denotes the dual map (Heisenberg picture of $\Phi$) defined via
\begin{equation}\label{}
  {\rm Tr}(A \Phi(B)) = {\rm Tr}(\Phi^\ddag(A) B) ,
\end{equation}
for any $A,B\in \LH$. If map $\Phi$ is trace-preserving (TP), i.e., it is a quantum channel \cite{holevo}, one has  $\mathcal{L}(\rho) = -i[H,\rho] + \Phi(\rho) - \rho$. Assuming $H=0$, the entire generator is uniquely defined by $\Phi$. Yet, such $\mathcal{L}$ is not purely dissipative. Indeed, one has
%\begin{widetext}
%\begin{equation}\label{}
  %\Phi(\rho) = \sum\limits_{m,n=0}^{N^2-1} \!\! %K_{mn} F_n\rho F^{\dagger}_m = K_{00} \rho + %\frac{1}{\sqrt{N}} \sum_{k=1}^{N^2-1} ( %F_n\rho + \rho F^{\dagger}_n) + %\sum\limits_{m,n=1}^{N^2-1} \!\! K_{mn} %F_n\rho F^{\dagger}_m
%\end{equation}
%\end{widetext}

\begin{multline}\label{}
  \Phi(\rho) = \sum\limits_{m,n=0}^{N^2-1} \!\! K_{mn} F_n\rho F^{\dagger}_m =
  \\
  K_{00} \rho + \frac{1}{\sqrt{N}} \sum_{k=1}^{N^2-1} ( F_n\rho + \rho F^{\dagger}_n) + \sum\limits_{m,n=1}^{N^2-1} \!\! K_{mn} F_n\rho F^{\dagger}_m
\end{multline}
and hence $\mathcal{L}$ can be represented by (\ref{L0}) with the residual Hamiltonian
\begin{equation}\label{}
  H_0 = \frac{i}{2\sqrt{N}} \sum_{l=1}^{N^2-1} ( K_{l0} F_l - \overline{K}_{l0} F_l^\dagger ) .
\end{equation}
Note, however, that in the large $N$ limit the above Hamiltonian vanishes and hence any purely dissipative random Lindbladian can be (for $N$ large enough) represented by a  completely positive map $\Phi$. In what follows we refer to
\begin{equation}\label{LL}
  \mathcal{L}(\rho) = \Phi(\rho) -
  \frac 12 \bigl( \Phi^\ddag(\oper) \rho + \rho\; \Phi^\ddag(\oper) \bigr),
\end{equation}
as the {\em Lindblad representation}. Again, in the Lindblad representation, the generator $\mathcal{L}$ is fully specified by an auxiliary CP map $\Phi$.

A random CP map $\Phi$ can be obtained from a random Choi state. A random Choi matrix can be sampled as Wishart matrix \cite{ZPNC11}, $\mathbf{C} = N\cdot GG^\dagger/\mathrm{Tr}GG^\dagger$, where $G$ is a complex $N^2 \times N^2$ square Ginibre matrix with i.i.d. complex Gaussian entries. This Hilbert-Schmidt ensemble induces the Lebesgue measure in the space of quantum states \cite{ZPNC11} and defines an ensemble of CP maps, through the standard reshuffling operation~\cite{res1,res2}, $\widehat{\Phi} = \mathbf{C}^R$ (note that the reshuffling operation is an involution). Since the trace preservation is not imposed on the map, this procedure is simpler than the one used to sample random channels presented in Refs.~\cite{BCSZ09,KNPPZ21}. 

For $N \gtrsim 100$, the sampling results in the same universal lemon-shaped distribution as the previously considered algorithm based on the GKS-representation; see Fig.~2c. In this case $N$ is limited only by the diagonalization cost and memory size, so it is possible to sample  Lindbladians for larger values of $N$~\cite{sample}.

\subsection{Random Matrix model}

Important spectral properties of random  Lindbladians can be studied with the help of a Random Matrix (RM)  model, derived as follows.

While the generation of random Lindbladians from random Choi states, Eq.~(\ref{LL}), is more convenient for numerical analysis, RM model easily follows from
the GKS-representation, Eq.~(\ref{L}). 
If the normalization of the trace of the Kossakowski matrix is relaxed from $\textup{Tr} K=N$ to $\< \textup{Tr} K\>=N$,  
%since the Choi matrix of the CP map $\Phi$ is positive definite, it can be represented  as a Wishart matrix $\mathbf{C}^\Phi = Z^{\dagger}Z$.  If the normalization condition $\textup{Tr} \mathbf{C}^\Phi = N$ is relaxed to $\< \textup{Tr} \mathbf{C}^\Phi\>=N$, 
the elements of $G$, Eq.~\eqref{eq:RandomKoss}, are i.i.d. Gaussian random variables with the variance
\begin{equation}
\<G_{ab}G^{\dagger}_{cd}\>=\frac{1}{N^3}\delta_{ad}\delta_{bc} . \label{eq:VarianceX}
\end{equation}
Introducing set of operators
$$   Y_k = \sum_{l=0}^{N^2-1} G_{kl} F_l,$$
where ${\rm Tr}(F_k F_l^\dagger) = \delta_{kl}$,  one can represent  map $\Phi$ in the following Kraus form:
\begin{equation}\label{}
  \Phi(\rho) = \sum_{k=0}^{N^2-1} Y_k \rho Y_k^\dagger.
\end{equation}
Note that
\begin{equation}\label{}
  \< \textup{Tr}\, Y_k Y_l^\dagger \> = \frac{1}{N}\, \delta_{kl} ,
\end{equation}
and entries of $Y_k$ are  independent, thus in the large $N$ limit eigenvalues of $Y_k$ cover uniformly the disk of radius $r$, where
\begin{equation}
r^2=\<\frac{1}{N}\textup{Tr}\, Y_k Y^{\dagger}_k\>=\frac{1}{N^2} .
\end{equation}
Matrices $Y_k$ allow us to rewrite the Lindblad super-operator as follows
\begin{equation}
\hat{\cL}=\hat{\Phi}-\oper\otimes \oper -\frac{1}{2}(X\otimes \oper + \oper \otimes \overline{X}), 
\label{eq:LindbladMatrixRepr}
\end{equation}
with
\begin{equation}
\widehat{\Phi} = \sum_{k=0}^{N^2-1} Y_k\otimes \overline{Y}_k,
\end{equation}
and the Hermitian operator
\begin{equation}\label{}
X = \Phi^{\ddag}(\oper) - \oper = - \oper + \sum_{k=0}^{N^2-1}Y_k^{\dagger}Y_k   .
\end{equation}
All $N^2$ eigenvalues of $Y_k\otimes\overline{Y}_k$ have the form  $\lambda_i\overline{\lambda}_j$ for $i,j=1,\ldots,N$, where $\lambda_{i}$ are eigenvalues of $Y_k$, thus their density is supported on a disk of radius $1/N^2$. The operator
 $\widehat{\Phi}$ is a sum of $N^2$ independent matrices $Y_k\otimes\overline{Y}_k$, and hence, according to the central limit theorem for non-Hermitian matrices~\cite{TAOVU}, its spectral density is uniform on the disk of radius $1/N$. As a consequence, in the large $N$ limit, $\widehat{\Phi}$ can be modeled as a Ginibre matrix with the spectral radius $1/N$. Note that
$$X = \sum_{k=0}^{N^2-1} X_k , $$
with $X_k=Y^{\dagger}_k Y_k - \frac{\oper}{N^2}$ being a shifted Wishart matrix. One finds that

$$  \<  {\rm Tr} X_k \> = 0 , $$
and for the variance

$$  \sigma^2 = \< \frac{1}{N} {\rm Tr} X_k^2 \> = \frac{1}{N^4} . $$
Now, since $X$ is a sum of $N^2$ such independent  matrices, then, according to the central limit theorem for Hermitian matrices \cite{Mehta}, its spectral density is the Wigner semicircle supported on $[-2/N,2/N]$,
\begin{equation}
\rho_X(x)=\frac{N^2}{2\pi}\sqrt{\frac{4}{N^2}-x^2}.
\end{equation}
The above reasoning correctly predicts the $1/N$ scaling and unit shift $\hat{{\cal L}}= -\mathcal{I} + \frac{1}{N}\hat{{\cal L}}'$ and justifies the following approximation
\cite{denysov2019}

%of $\hat{{\cal L}}'$
\begin{equation}\hat{{\cal L}}'\approx G_R-(C\otimes \oper+\oper \otimes \overline{C})=G_R-B, \label{eq:SMmodel}
\end{equation}
where $B=B^{\dagger}$ and the spectral density of $C$ is the Wigner semicircle on $[-1,1]$
\begin{equation}
\rho_C(x)=\frac{2}{\pi}\sqrt{1-x^2}.
\end{equation}
While the matrix representation of $\widehat{\mathcal{L}}$ is not real, Eq.~\eqref{TILDE} provides another representation, 
%$\widetilde{\mathcal{L}}$ (cf. (\ref{TILDE})) 
which is real.
% and has exactly the same spectrum as  $\widehat{\mathcal{L}}$.  
%This means that there exists a basis in which matrix elements of the Lindblad operator are real.
Therefore, one can take $G_R$ as a real Ginibre matrix
and $C$ as a symmetric GOE matrix so that $\overline{C}=C$.
Spectral distribution on the complex plane for a wide class of random matrix models 
composed of a non-Hermitian, Ginibre part plus an independent Hermitian part,  was investigated in Ref.~\cite{Kh96}.

\subsection{Spectral densities of random Lindblad generators}

\begin{figure*}[t]
\begin{center}
	\includegraphics[angle=0,width=0.99\textwidth]{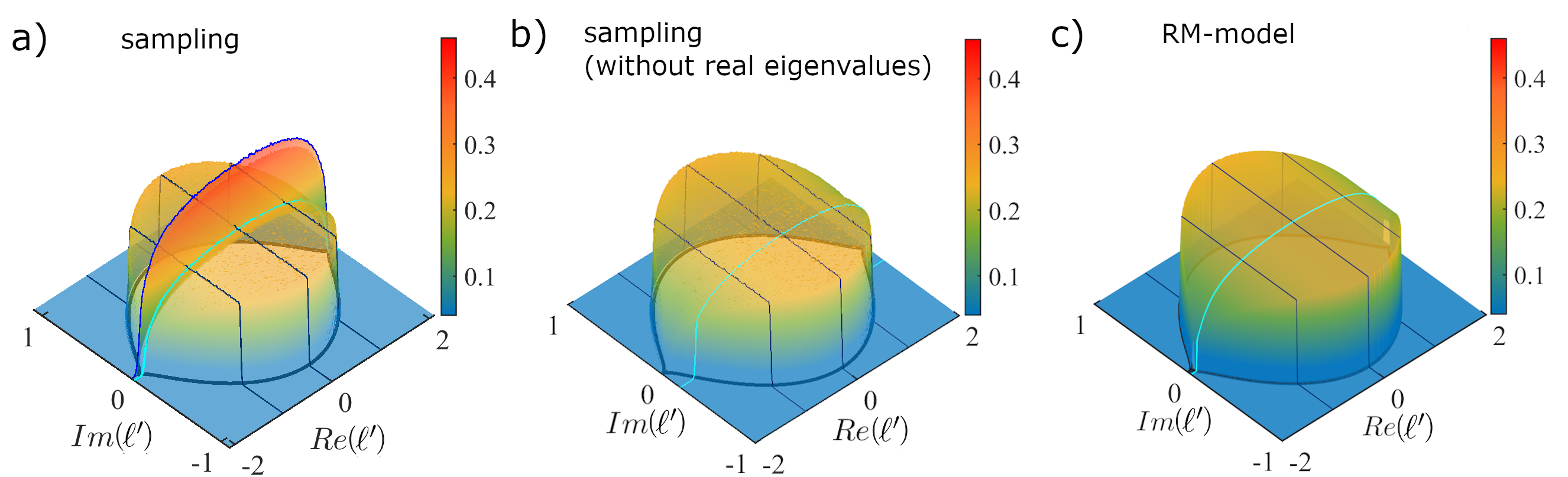}
%\end{center}	
\caption{Probability density functions
$P[\mathrm{Re}(\ell'),\mathrm{Im}(\ell')]$
of the rescaled eigenvalues,	$\ell'= N (\ell +1)$,
of random Lindblad operators  ${\cal L}$ for $N=100$ (a-b) and Random Matrix model (c). The cross-section indicated with lines are addressed with Fig.~3. Densities on (a-b) are sampled  with $10^3$ realizations. Bright contours on the complex plane correspond to the spectral boundary,  Eqs.~(\ref{eq:BorderlineFinalM}-\ref{eq:contour2}) Note that eigenvalue $\ell_1 = 0$ is excluded.
\label{fig:2}
}
\end{center}	
\end{figure*}

Spectral properties of the random matrix model  Eq.~(\ref{eq:SMmodel}), 
designed to mimic behaviour of  random Lindblad operators (\ref{eq:LindbladMatrixRepr}),
can be studied with the help of analytical tools of free probability~\cite{S2,S3,S4,S5,S6,S7}. Detailed derivation is presented in Appendix A. Here we outline the main results.

Since the matrix model is real, eigenvalues are either real or come in complex conjugate pairs. The eigenvalue density $P[\textup{Re}(\ell'),\textup{Im}(\ell')]$, which we also denote as $\rho(z,\zb)$ following the RMT literature, consists of the density of complex eigenvalues $\rho_c(z,\zb)$ and the density of purely real eigenvalues $\rho_r(x)$. Typically in RMT, the number of real eigenvalues grows proportionally to the square root of the matrix size~\cite{Edelman}, thus the latter density is negligible in the large $N$ limit. Nevertheless, their presence markedly affects the spectra of finite matrices; see Fig.~3a.

The boundary of the lemon-like bulk of complex eigenvalues is characterized by the solution of the following equation
\begin{equation}
\textup{Im}\left[ z+G(z)\right]=0,
\label{eq:BorderlineFinalM}
\end{equation}
with
\begin{multline}
G(z)=  \\
2z-\frac{2z}{3\pi}\left[(4+z^2)E\left(\frac{4}{z^2}\right)+(4-z^2)K\left(\frac{4}{z^2}\right)\right],
\label{eq:contour2}
\end{multline}
where $K(z)$ and $E(z)$ are the complete elliptic integrals of the first and second kind, respectively. Results of a sampling for $N \geq 50$ are in a perfect agreement with numerical solutions of Eq.~\eqref{eq:contour2}; see Fig.~2.

Density of complex eigenvalues can be expressed in a rather complicated form that involves solutions of Eq.~\eqref{eq:BorderlineFinalM}. 
%and the function $G(z)$. 
Its main feature is that the distribution is constant %along lines $\mathrm{Re}(\ell') = \mathrm{const}$ in the interior of the lemon.
in the imaginary direction inside the lemon.
This form, given in Appendix A [see Eq.~\eqref{eq:DensityComplex} in there], can be numerically evaluated and compared with the sampled spectral distribution. We find a perfect agreement between the two results, see Figs.~3(a-c).

\begin{figure}[b]
\begin{center}
	\includegraphics[angle=0,width=0.49\textwidth]{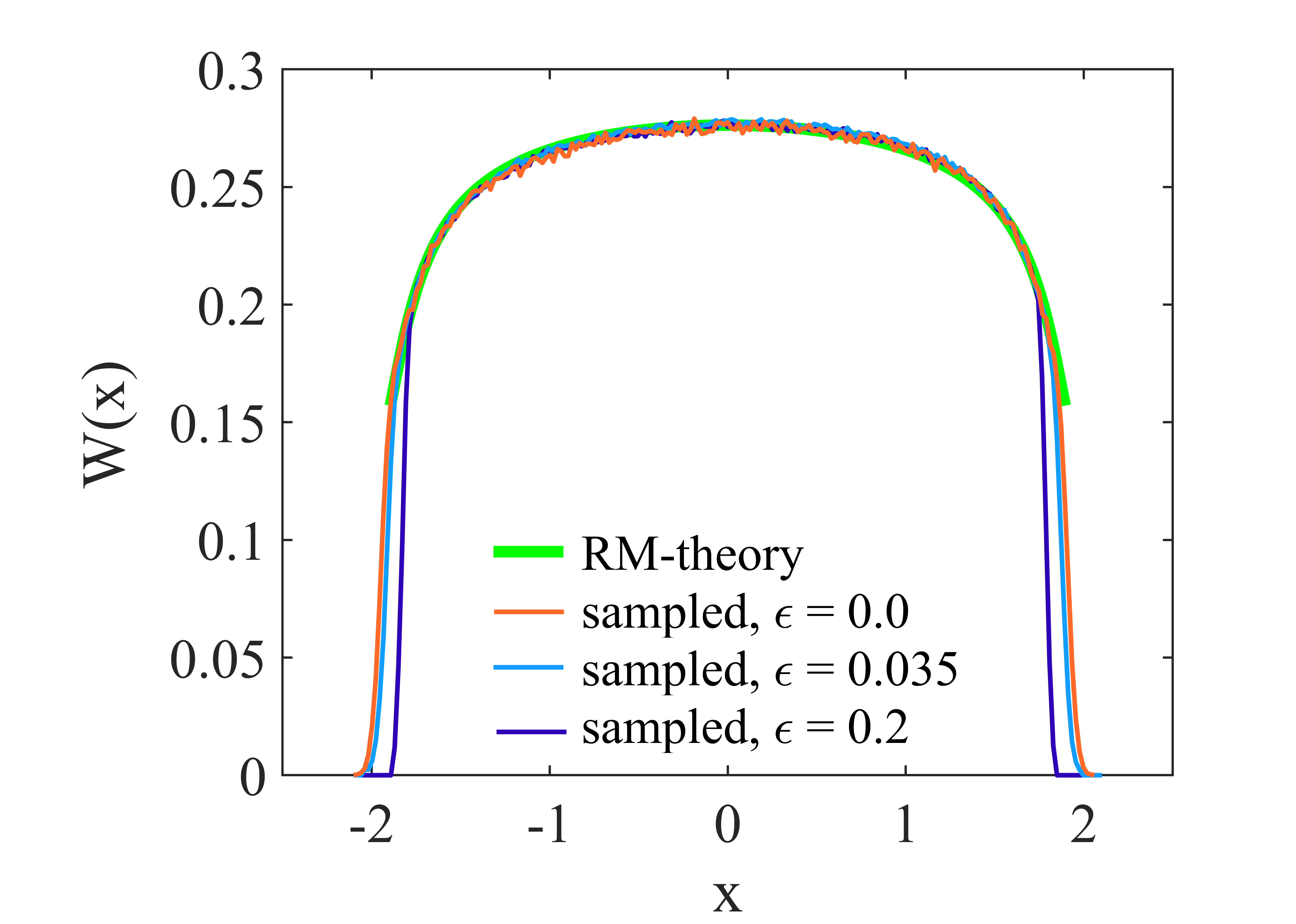}
%\end{center}	
\caption{Sampled density of real eigenvalues (red) juxtaposed with the square root of the theoretical density of complex eigenvalues evaluated at the real line (green), Eq.~\eqref{eq:RealDensity}.  Blue and violet lines present square root of the complex density evaluated along the lines $z=x+i\epsilon$ with $\epsilon = 0.035$ and $0.2$, depicted on Fig.~3(a-b). 
%Dashed and solid lines coincide in the %central region because the density is %constant along the imaginary direction.
}
\end{center}	
	\label{fig:3}
\end{figure}

As we mentioned earlier, since the Lindblad operator has a real representation, the spectral density near the real axis deserves special interest for large but finite matrices.
There is a concentration of eigenvalues at the real line $\mathrm{Im}(z)=0$ (see Fig.~3a) which repel complex eigenvalues causing their depletion for small, but non-zero values of $\mathrm{Im}(z)$.
%[bright cyan lines on Figs.~2(a,c)].

Numerical evaluation of the density involves the difference of the Green's function evaluated at the opposite sides of the spectral boundary, $G(z)-G(\bar{z})$, and the division by its width, $z-\bar{z}$. This procedure becomes numerically unstable close to the tip of the lemon, where $G(z)$ needs to be evaluated at the opposite sides of its branch cut in the vicinity of the branch point. Being aware of this issue, we truncate the curve before the numerical instability region; see Figs.~3(b-c).

While free probability provides tools for analyzing complex eigenvalues, it does not provide a prescription for the density of real eigenvalues. However, it turns out that the density of real eigenvalues can be obtained directly from the asymptotic density of complex eigenvalues, which is given by Eq.~\eqref{eq:DensityComplex}. More specifically,
\begin{equation}
    \rho_r(x) \sim \sqrt{\rho_c(x,\epsilon)},
    \label{eq:RealDensity}
\end{equation}
that is, the density of real eigenvalues is proportional to the square root of the density of complex eigenvalues $z = x + i\epsilon$ evaluated along the real direction. This remarkable relation has a geometric origin and it stems from the Jacobians of appropriate change of variables~\cite{WTReals}. This prediction is perfectly verified by the results of the numerical sampling; see Fig.~4.

%the density of real eigenvalues is not accessible within free probability, based on the existing RMT results~\cite{Edelman,Simm1,ForresterIpsen,Simm2}, we conjecture that this density (inside the lemon) is proportional to the square root of the  density along lines  $\mathrm{Im}(\ell')=\epsilon$, $\vert \epsilon \vert \neq 0$ i.e., along the line outside the eigenvalue concentration/depletion region near $\mathrm{Im}(\ell')=0$ [bright cyan lines on Figs.~2(a,c)].
%evaluate on the real axis, $\rho_r(x) \sim %\sqrt{\rho_c(z=\zb=x)}$. 

\section{Spectra of random Kolmogorov operators}\label{sec:2c}

Consider now a classical Markov semigroup, determined by a family of stochastic matrices (classical dynamical map),
 % $\mathsf{T}$
 %% attention: nottaion change   T--> S  for stochastic matrix
 % to be consistent with new introduction !
$S_t = e^{t\cal{K}}$, where  $N \times N$ matrix $\cal{K}$ (henceforth called `Kolmogorov generator') satisfies \cite{S13,classic}:

$$   {\cal{K}}_{ij} \geq 0 \ , \ (i \neq j) \ , \ \ \sum_{i=1}^N {\cal{K}}_{ij} = 0 \ , \ (j=1,\ldots,N) . $$

The spectrum of random stochastic matrix consists of eigenvalues $\chi_1=1$ while the rest of eigenvalues fills the characteristic Girko disk of radius $1/\sqrt{N}$ \cite{S14}. Any Kolmogorov generator may be represented in terms of a real matrix $M_{ij} \geq 0$ via ${\cal{K}}_{ij} = M_{ij} - \delta_{ij} m_j$, with $m_j = \sum_i M_{ij}$. The diagonal elements $M_{ii}$ are not essential, since they cancel out.

We assume that  elements $M_{ij}>0$ are i.i.d. sampled from distribution with first two moments $m_1=\int xp(x)dx=\frac{\mu}{N}$ and $\int (x-m_1)^2p(x)dx=\frac{\sigma^2}{N}$. The underlying distribution should not play essential role and one could expect, similar to the quantum case, a high degree of universality. E.g., a random $N \times N$ stochastic matrix 
% $\mathsf{T}$
$M_{ij}$ may consists of $N$ random %probability 
positive vectors, $M_{ij} = |z_{ij}|^2$
%$M_{ij} = |z_{ij}|^2/ {\sum_i |z_{ij}|^2}$, 
where $z_{ij}$ are i.i.d. Gaussian complex variables. 

We write ${\cal{K}}=M-J$, where $J_{ij}=\delta_{ij}\sum_{k}M_{kj}$. Having this decomposition, we now inspect closer these two matrices. The elements of the matrix $M$ are i.i.d., thus, according to the circular law, its spectral density is uniform on the disk of radius $\sigma$, located at 0. There is also a single eigenvalue located around $\mu$. The elements of $J$ are sums of independent random variables, so the diagonal elements in the large $N$ limit are Gaussian with mean $\mu$ and variance $\sigma^2$.

\begin{figure} [t]
\label{FIG-class}
\begin{center}
\includegraphics[width=0.4\textwidth]{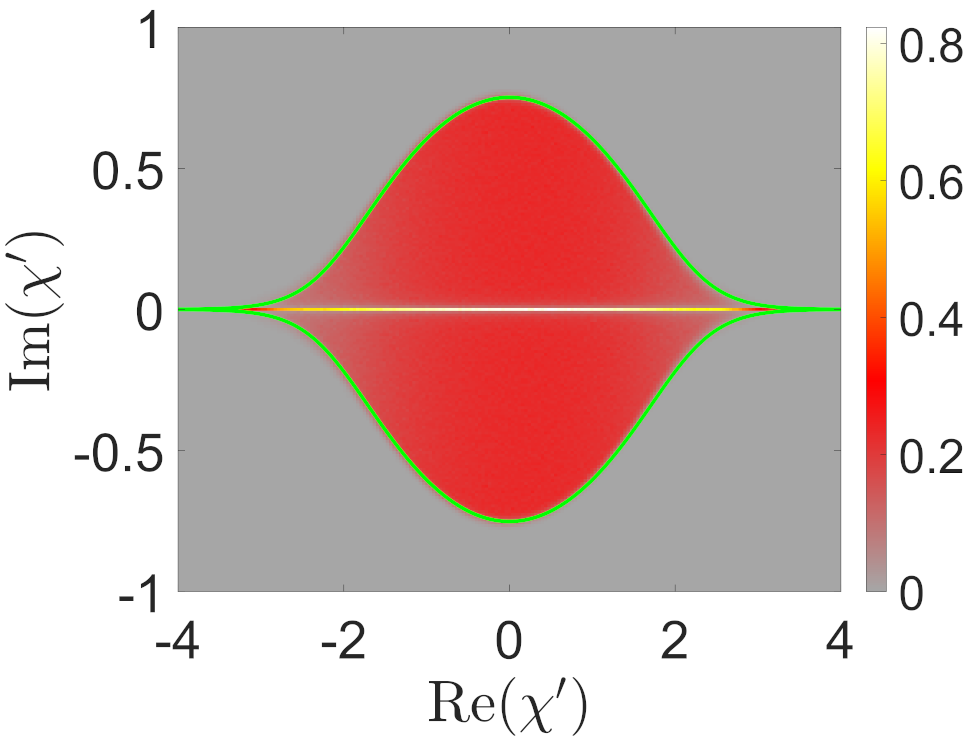}
\end{center}
\caption{Probability density functions
$P[\mathrm{Re}(\chi'),\mathrm{Im}(\chi')]$
of the rescaled eigenvalues,	$\chi'= \sqrt{N} (\chi'+1)$,
of random Kolmogorov operator  $\cal{K}$ for $N=2000$. Bright green contour is the spectral border, Eqs.~(\ref{eq:BorderlineFinalM},\ref{eq:contour3}). Total number of samples is $10^3$.
\label{fig:5} 
}
\end{figure}
We can therefore write
\begin{equation}
{\cal K}=-\mu \oper+\sigma(G_R + D)
\end{equation}
where $G_R$ is real Ginibre with radius 1 and the eigenvalues of $D$ follow the normal distribution. In the same spirit as with the Lindblad generators, we decomposed the Kolmogorov generator into a shift (by $\mu$) and a scaling (by $\sigma)$. The non-trivial part, ${\cal K}'=G_R+D$ is given by
 a free convolution of a Girko disk and a Hermitian Gaussian distribution.
 
\begin{figure*}[t]
\begin{center}
\includegraphics[angle=0,width=0.99\textwidth]{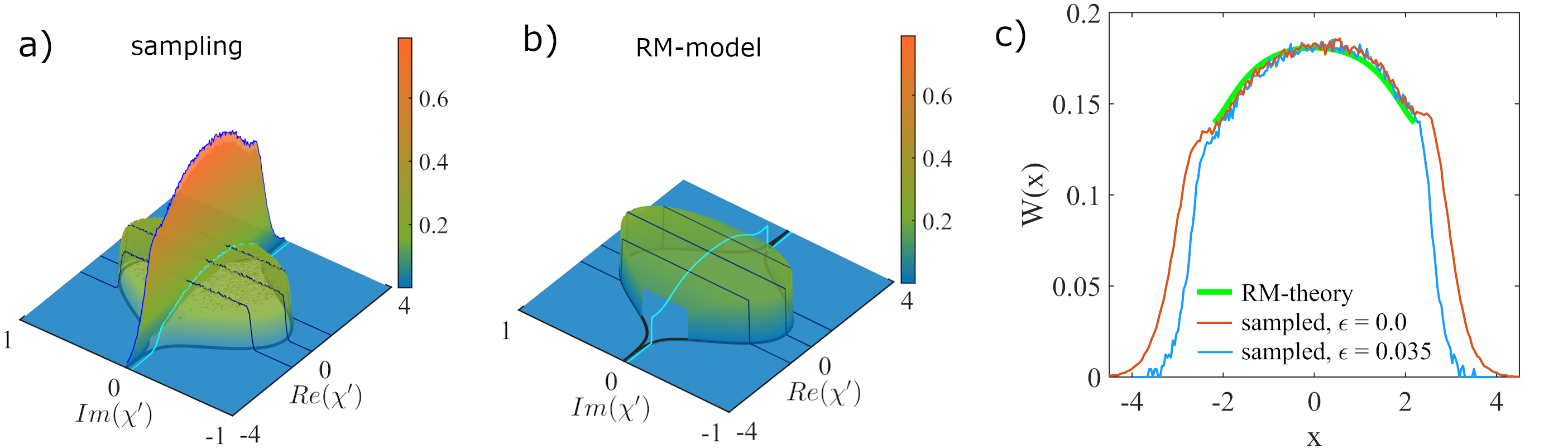}
%\end{center}	
\caption{Probability density functions
$P[\mathrm{Re}(\chi'),\mathrm{Im}(\chi')]$
of the rescaled eigenvalues,	$\chi'= \sqrt{N} (\chi+1)$,
of random Kolmogorov operators  $\cal{K}$ for $N=2000$ (a) and Random Matrix model (b).   Bright contours on the complex plane correspond to the spectral boundary, Eqs.~(\ref{eq:BorderlineFinalM},\ref{eq:contour3}). Densities on (a) are sampled  with $10^3$ realizations. The analytically obtained  distribution is truncated near the cusp regions where the numerical evaluation experienced instability.  (c) Juxtaposition of the density of the real eigenvalues (red) with the square root of the density of complex eigenvalues (green). Blue line is the square root of the eigenvalue density $W(x)=P[x,\epsilon]$, $\epsilon = 0.035$.}
\end{center}	
	\label{fig:5}
\end{figure*}

We notice that, in the full analogy with random Lindbladians, ${\cal {K}}'$ is a sum of a real Ginibre matrix and a Hermitian matrix, thus the same formalism of free probability described in Appendix A applies here. The only difference is that in Eq. \eqref{eq:Greens} one needs to use the normal distribution of eigenvalues as $\rho_B$.  
We get
%\begin{widetext}
%\begin{eqnarray}
%G_{K}(z)=\frac{1}{\sqrt{2\pi}}\int_{-\infty}^{\infty} \frac{e^{-x^2/2}}{z-x}dx =
%\sqrt{\frac{\pi}{2}}e^{-z^2/2}\left(\textup{Erfi}\left(\frac{z}{\sqrt{2}}\right)-i \textup{sgn} (\textup{Im} z)\right),~~~~
%\label{eq:contour3}
%\end{eqnarray}
%\end{widetext}

\begin{multline}
G_{{\cal K}'}(z)=\frac{1}{\sqrt{2\pi}}\int_{-\infty}^{\infty} \frac{e^{-x^2/2}}{z-x}dx = \\ 
\sqrt{\frac{\pi}{2}}e^{-z^2/2}\left(\textup{Erfi}\left(\frac{z}{\sqrt{2}}\right)-i \textup{sgn} (\textup{Im} z)\right),~~~~
\label{eq:contour3}
\end{multline}
where $\textup{Erfi}(z)=-i\textup{Erf}(iz)$ and $\textup{Erf}(z)=\frac{2}{\sqrt{\pi}}\int_0^z e^{-t^2}dt$ is the error function. The boundary of ${\cal K}'$ can be calculated with Eq.~\eqref{eq:BorderlineFinalM}
in which we now substitute $G_{{\cal {K}}'}(z)$.

In the classical regime the  support of the spectra of random Kolmogorov generators
has a spindle-like shape, as it has cusps along the real axis that are more pronounced 
than the cusps of the lemon-like contour characteristic to the spectra of  random Lindbladians.
 Within the RMT framework, this is a result of the free convolution of the Girko disc with the Gaussian distribution which does not have a compact support, as opposed to the Meijer G-function in the quantum case (see Fig.~\ref{Fig:SpecB} in Appendix A). The comparison with the results of a numerical sampling for $N = 2000$ is presented in Fig.~5.

It is noteworthy that our result is in a full agreement with the results obtained by Timm in his first work on the subject~\cite{timm}
and recently extended in Ref.~\cite{LT21}. 
Spectra of random ensembles of Kolmogorov operatorss have also been studied with the apparatus of free probability in Ref.~\cite{S15}, but in a different context since the operators were sampled by using  random graphs.
Nevertheless, Fig.~1(bottom) in Ref.~\cite{S15} reveals the spectral distribution
resembling the spindle presented in our Fig.~5.

Similar to the quantum case, the spectral density can  be evaluated with free probability. Figure~6b presents  analytical results which we compare with sampled probability density functions (Fig.~6a). Similar to the case of Lindblad operators, the numerical procedure becomes unstable as the width of the spectrum decreases, so we truncate the analytical distribution on Fig.~6b near the cusps.

To evaluate the density of real eigenvalues, we again apply Eq.~\eqref{eq:RealDensity} stating that this density is proportional to the square root of the density along lines  $\mathrm{Im}(\chi')=\epsilon$, $\vert \epsilon \vert > 0$.
%evaluate on the real axis, $\rho_r(x) \sim %\sqrt{\rho_c(z=\zb=x)}$. 
This conjecture is confirmed by the results of the numerical sampling; see Fig.~6c.

\section{Superdecoherence: from Lindblad to Kolmogorov operators}
\label{sec:3}

Now we demonstrate how the two  types of generators, quantum and classical, can be related -- in a continuous way -- by superdecoherence. In particular, one could  obtain an ensemble of random Kolmogorov operators by subjecting an ensemble of random Lindblad operators to complete decoherence.

Consider Lindbladian $\mathcal{L}$ defined in Eq.~(\ref{LL}) in terms of a completely positive map $\Phi$. In its turn, map $\Phi$ ban be represented in terms of its Choi matrix $\mathbf{C}$ as: 
\begin{equation}\label{}
  \Phi(\rho) = \sum_{i,j,k,l=1}^N \mathbf{C}_{ij,kl} |i \rangle\langle j|\, \rho\, |l \rangle\langle k| .
\end{equation}
Now, let us perform a superdecoherence with parameter $p \in [0,1]$ via the following Hadamard product
\begin{equation}\label{S-DEC}
  \mathbf{C} \to \mathbf{C}^{(p)} = \widetilde{\Delta}^{(p)} \circ \mathbf{C} ,
\end{equation}
where $\Delta^{(p)}$ is defined as 
\begin{equation}\label{}
  \widetilde{\Delta}^{(p)}_{ij,kl} = \begin{cases}
      1,  \mathrm{if}\ (ij)=(kl), \\
      p ,  \mathrm{otherwise},
    \end{cases}
\end{equation}
One has therefore

\begin{equation}
\mathbf{C}^{(p)}_{ij;kl} = \begin{cases}
      \mathbf{C}_{ij;kl},  \mathrm{if}\ (ij)=(kl), \\
      p\cdot \mathbf{C}_{ij;kl},  \mathrm{otherwise} ,
    \end{cases}
\label{eq:decoh}
\end{equation}
that is, the off-diagonal matrix elements of $\mathbf{C}$ are suppressed by a factor $p$ (superdecohered). Evidently, $\mathbf{C}^{(p)} \geqslant 0$. Therefore, it corresponds to a CP map $\Phi^{(p)}$ which can be used to construct a new Lindbladian via

\begin{figure} [b]
\label{FIG-class}
\begin{center}
\includegraphics[width=0.4\textwidth]{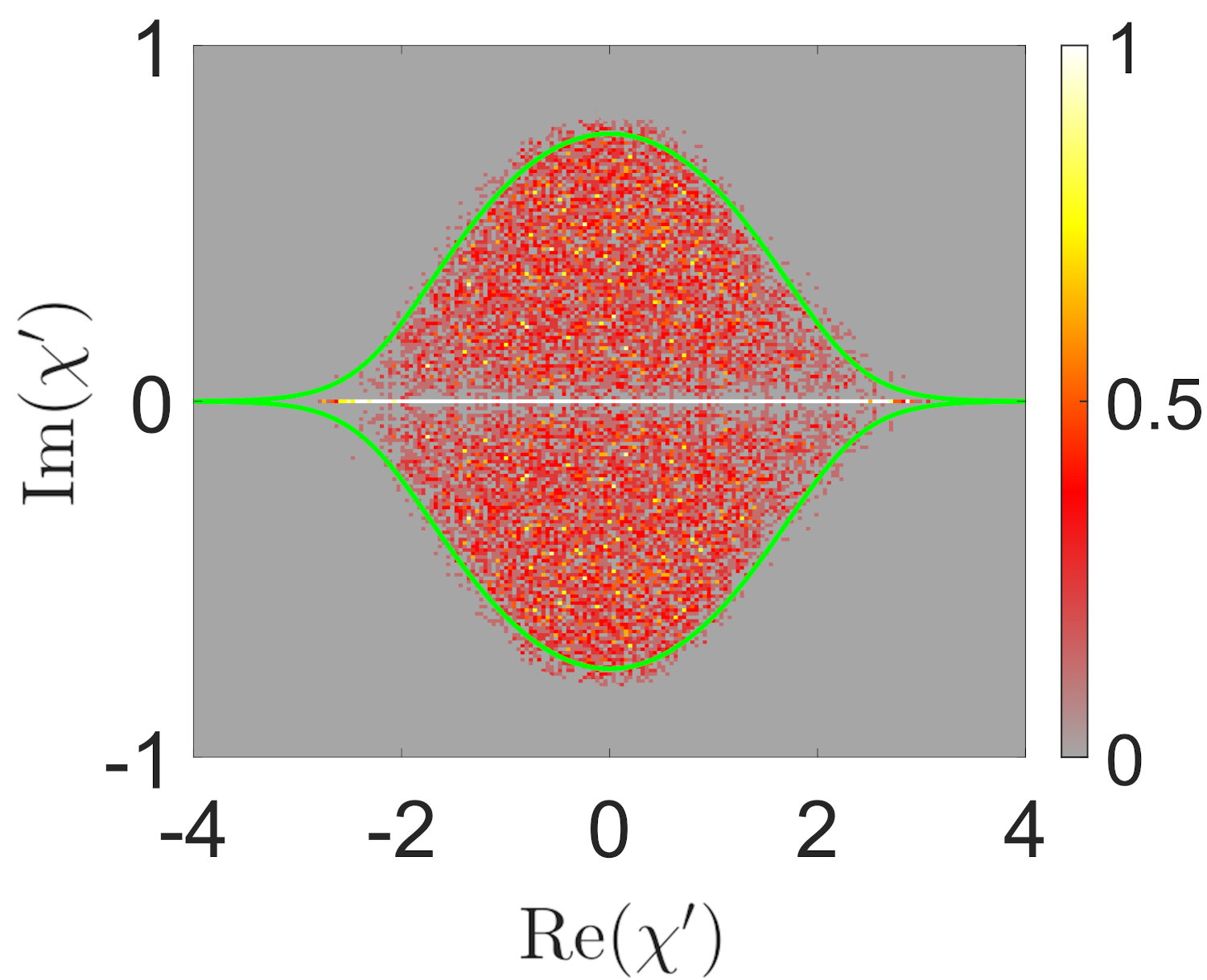}
\end{center}
\caption{Probability density function
	$P[\mathrm{Re}(\ell'),\mathrm{Im}(\ell')]$
	of the rescaled eigenvalues,	$\ell'= N^{\frac{3}{2}} (\ell +1)$,
	from the spectrum of ${\cal L}^{p=0}$, Eq.~\eqref{eq:decoh}. The distribution for $N=200$ was sampled with $10^2$  realizations. Bright green contour is  the spectral boundary of random Kolmogorov generators, Eqs.~(\ref{eq:BorderlineFinalM},\ref{eq:contour3}). $N-1$ real $N$-fold degenerated eigenvalues, corresponding to the decoupled evolution of the off-diagonal elements of the density matrix, contribute to the  white line on the real axis.}
\label{fig:6}
\end{figure}

\begin{equation}\label{}
  \mathcal{L}^{(p)}(\rho) = \Phi^{(p)}(\rho) - \frac 12 \{ \Phi^{(p)\ddag}(\oper),\rho\} .
\end{equation}
In the limit $p \rightarrow 0$, the evolution of the diagonal elements of density operator $\rho$ under the action of $\mathcal{L}^{(p)}$ decouples from the evolution of the off-diagonal elements and for diagonal elements we thus obtain Kolmogorov generator $\cal{K}$.
The evolution of the off-diagonal elements is governed by a generator with pure real negative $N$-fold degenerate spectrum. We observe that the spectrum of random generator $\mathcal{L}^{(p=0)}$ recovers the universal structure of the spectra of random Kolmogorov generator if the former is rescaled by $N^{\frac{3}{2}}$. Recall that the universal structure of quantum Lindbladian $\mathcal{L}^{(p=1)}$ requires the scaling by $N$. Results of the sampling for $N=200$ are shown in Fig.~7. Interestingly, if %one switches on 
the superdecoherence with  $p \lesssim 0.5$ is switched on, we still observe the universal `lemon' structure provided it scaled with $N/p$; see Fig.~8.

\begin{figure}[t]
\begin{center}
\includegraphics[width=0.49\textwidth]{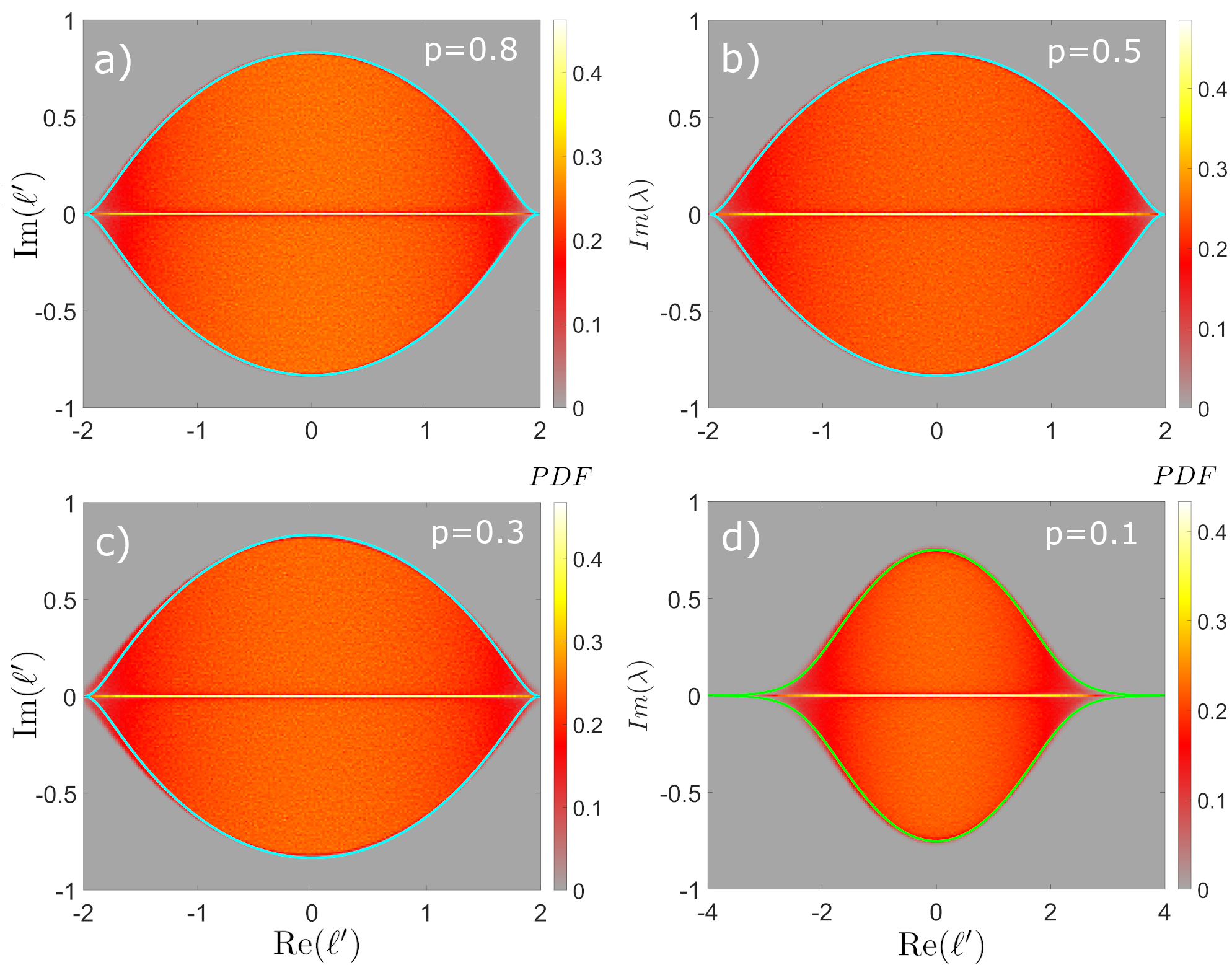}
\end{center}
\caption{Probability density function
$P[\mathrm{Re}(\ell'),\mathrm{Im}(\ell')]$
of the rescaled eigenvalues,	$\ell'= w(N)(\ell +1)$, of ${\cal L}^{p}$, Eq.~\eqref{eq:decoh}, for different values of $p$, $N=100$. For $p = 0.8$, $0.5$, $0.3$,  distributions were scaled with $w(N) = N/p$ while for $p=0.1$ the scaling factor $w(N) = N^{\frac{3}{2}}$. Bright (cyan) contours (a-c) corresponds to the spectral boundary of random Lindblad generators, Eqs.~(\ref{eq:BorderlineFinalM},\ref{eq:contour2}), while bright (green) contour (d) corresponds to spectral boundary for random Kolmogorov generators, Eqs.~(\ref{eq:BorderlineFinalM},\ref{eq:contour3}).
Densities  were sampled with $10^3$ realizations each.
\label{fig:7}
}
\end{figure}

These two principally different scalings, for $p$ values close to one and close to zero, provide an evidence that there must be a sort of a phase transition. To inspect the continuous transition, from $p=1$ (quantum) to $p=0$ (classical), we need
to quantify the distance from the actual spectral distribution to the two limiting shapes, the quantum 'lemon' and classical 'spindle'. The immediate choice would be one the standards metrics used to quantify  difference between two probability density distributions, like Kullback–Leibler divergence or total variation distance \cite{distance}. However, in the realm of two-dimensional, statistically sampled, distributions (histograms), characterized by high irregularities, these standard theoretical tools perform badly.
A more reasonable choice in this situation is to use spectral boundaries which remain -- as we observe -- sharp for all values of $p$ if $N \gtrsim 50$.

To quantify the difference between the two contours,  we use the Jaccard distance \cite{jacck1}, a standard tool to gauge the similarity and diversity of two geometric sets. In our case the index reduces to the normalized (by the total joint area) overlapping area enclosed by two contours; see Appendix B for more details. In the case of planar geometric objects, it is also known as 'Intersection over Union' ($\mathrm{IoU}$) \cite{jacck1,jacck2}. In words, it is the overlap area  divided by the join area of two figures. The distance between the two contours $A$ and $B$ can be defined as $d(A,B)=1-\mathrm{IoU}(A,B)$.
Before calculating $\mathrm{IoU}$ for the sampled distribution and one of the contours, we scale the first with factor $w(N)=N/p$ (to compare with the lemon contour) or $w(N)=N^{3/2}$ (spindle contour). The two corresponding $\mathrm{IoU}$s are shown on Figure 9a as  functions of $N$.
%Figure 8 presents IoU as a function of $p$ %for different values of $N$.

\begin{figure}[b]
\begin{center}
\includegraphics[width=0.49\textwidth]{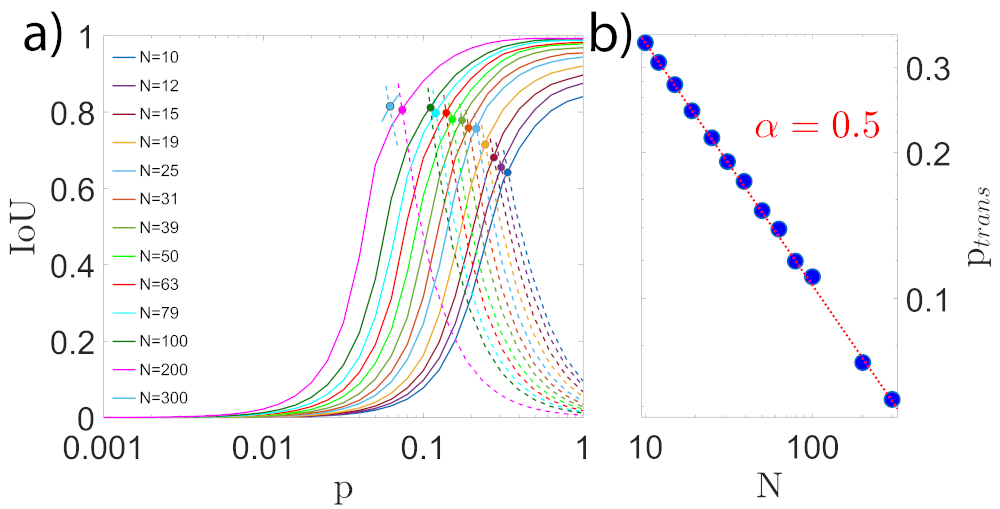}
\end{center}
\caption{Quantum-to-classical transition. (a)
Intersection over Union of the sampled distribution with the lemon-like contour, Eqs.~(\ref{eq:BorderlineFinalM}-\ref{eq:contour2}), (solid lines) and spindle-like contour, Eqs.~(\ref{eq:BorderlineFinalM},\ref{eq:contour3}), (dashed line) 
as functions of $p$ and $N$. When IoU equals one, the two spectral contours are identical.
The transition point $p_{\mathrm{tr}}(N)$ is defined as the value of $p$
at which  $\mathrm{IoU}$s with both limiting contours are equal. (b) Scaling of the transition point with $N$. Line corresponds to  $p(N) = 1/\sqrt{N}$.
\label{fig:8}
}
\end{figure}

\begin{figure*} [t]
\begin{center}
\includegraphics[width=0.99\textwidth]{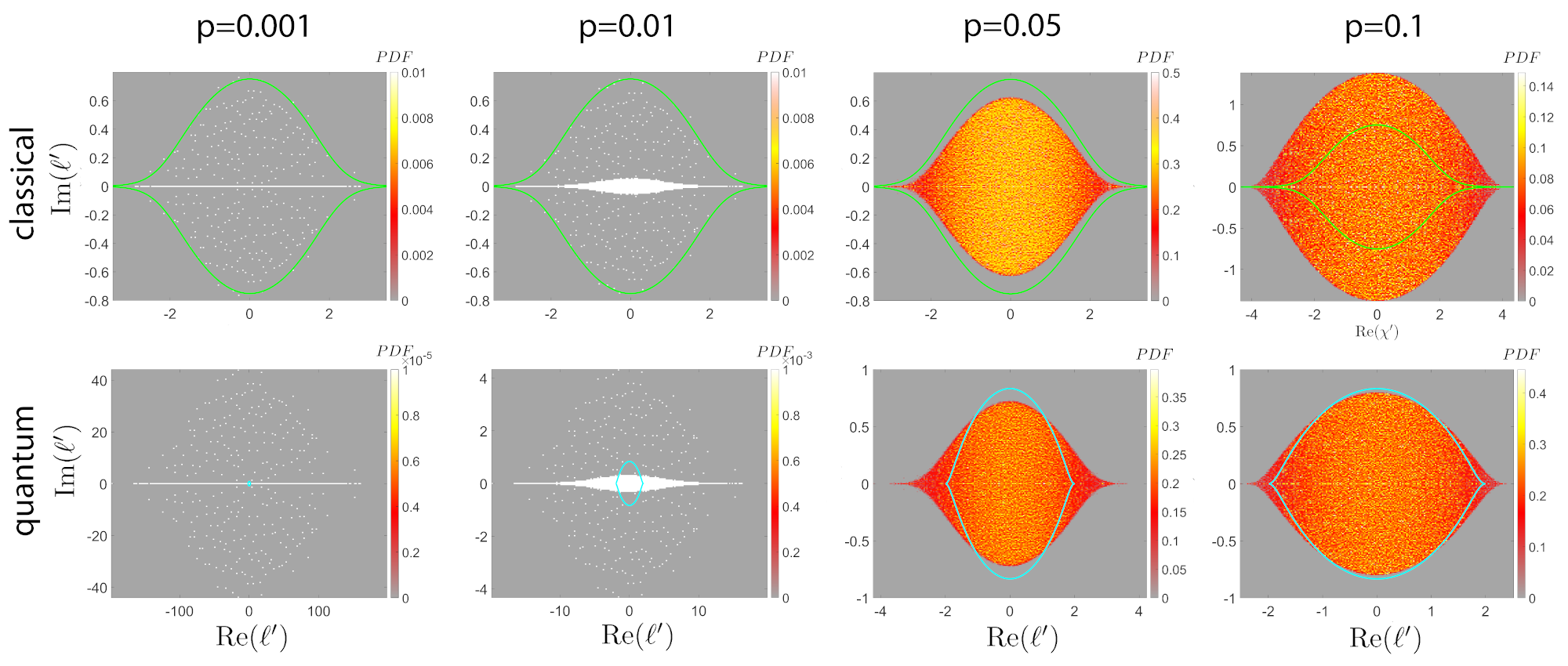}
\end{center}
\caption{Probability density function
$P[\mathrm{Re}(\ell'),\mathrm{Im}(\ell')]$
of the rescaled eigenvalues,	$\ell'= w(N)(\ell +1)$, of ${\cal L}^{p}$, Eq.~\eqref{eq:decoh}, for different values of $p$, obtained from a single sample for $N=300$.
In the upper row the distribution is scaled with $w(N) = N/p$,  while in the lower one it is scaled with $w(N) = N^{\frac{3}{2}}$. Bright (cyan) contours in the top row  is the spectral boundary of random Lindblad generators, Eqs.~(\ref{eq:BorderlineFinalM}-\ref{eq:contour2}), while bright (green) contour in the bottom row is the spectral boundary of random Kolmogorov generators, Eqs.~(\ref{eq:BorderlineFinalM},\ref{eq:contour3}).}
\end{figure*}

%Taking into account the two limiting  scalings, i.e., $N/p$ %and $N^{3/2}$, we  calculate  IoU of the corresponding %spectral distribution with lemon and the spindle as  reference %sets, and plot them as a function of $p$; see Fig.~8a. 

Next, we define the value $p_{\mathrm{tr}}(N)$, at which the two IoU curves intersect each other,  as the transition point. We find that, remarkably,  $p_{tr}(N)$ follows near exactly the dependence $N^{-\frac{1}{2}}$; see  Fig.~9b. 
Already starting $N=100$, the values $p_{\mathrm{tr}}$ obtained with the sampling  based on GKS- and Lindblad-representations  are identical (within the numerical accuracy) and so we use the last one to be able to sample for $N > 100$.  Even though with the Lindblad-representation we are able to sample Lindbladians for $N=300$, one such sample takes almost $18$ hours of computations on several cluster nodes. So, for every value of $p$, presented on Fig.~9 (five altogether), we take one sample for $N=300$. However, it is enough to determine the  boundary of the corresponding spectrum with high accuracy; see Fig.~10. Note that in the opposite limit of small $N$'s, the distance from the lemon is substantially nonzero even at the limit $p=0$. That is because for small systems the corresponding spectral distributions deviate from the asymptotic universal density.

The transition point can be  understood with the following  observation: Starting from  $p=1$, spectral distributions scale  $\sim \frac{N}{p}$, while on the 'classical' end, near $p=0$, the spindle does not depend on $p$ at all, and scaling goes as $N^{\frac{3}{2}}$. The two scalings meet at the point ${N}/{p_{\mathrm{tr}}} = N^{\frac{3}{2}}$. From this follows $p_{\mathrm{tr}} = N^{-\frac{1}{2}}$.
A similar scaling behavior describes the quantum-to-classical transition induced by superdecoherence on  the level of  quantum channels,  as was reported recently in Ref.~\cite{KNPPZ21}.

The classical-to-quantum transition is sharp already from $N=50$; note the logarithmic scale of the $p$-axis on Fig.~9a. It becomes sharper upon the further increase of $N$ and therefore bears features of a phase-transition at $p=0$~\cite{stanley}. While a more detailed investigation of the transition is an interesting task, it goes beyond the scope of this work and therefore is reserved for further studies.

There is another feature that cannot be captured with IoU, i.e., by using the spectral contours only. Upon the decrease of the value of $p$, the spectrum first is acquiring the  shape of the classical spindle-like contour, without a visible separation between the two classes of eigenvalues, corresponding to the evolution of diagonal and off-diagonal elements of the density matrix. The condensation of the latter on the real axis happens at the very last stage, see Fig.~10.

\section{Supercoherification: from Kolmogorov to Lindblad operators}
\label{sec:4}

Decoherence shapes quantum state, a density operator $\rho$, expressed in a certain basis, into a diagonal matrix ${\rm diag}(\rho)$ representing a  classical probability vector $\mathbf{q}=(q_1,\ldots,q_N)$. Given a  classical state $\mathbf{q}$ one can ask the following question: What are quantum states that can be decohered into it?

The answer could be obtained with  a procedure called {\it coherification}~\cite{COH}. Intuitively, it could be think as the inverse of decoherence defined with Eq.~(\ref{11}). Note, that any classical state $\mathbf{q}$ can be coherified into a pure quantum state $\mathbf{Q} = |\psi\rangle \langle \psi|$, where $|\psi\rangle = (\sqrt{q_1}e^{i\alpha_1},\ldots,\sqrt{q_N}e^{i\alpha_N})$, with arbitrary phases $\alpha_k$. Clearly, it is an extremal scenario and hence if we partially decohere $|\psi\rangle \langle \psi|$, then we can interpolate between a classical state $\mathbf{q}$ and a pure quantum state $\mathbf{Q}$. Let $B >0$ be an arbitrary (strictly) positive operator. Define a diagonal matrix 

$$   D_B = {\rm diag}(1/\sqrt{B_{11}},\ldots,1/\sqrt{B_{NN}}), $$
and let $\Delta = D_B B D_B$. One has $\Delta > 0$ and $\Delta_{kk}=1$ for $k=1,\ldots,N$. Finally, define a density operator as $ \rho_{kl}  =  \Delta_{kl} Q_{kl}$. Note, that $\rho_{kk} = q_k$, that is, we constructed a partial coherification of the original classical state $\mathbb{q}$. It should be stressed that any quantum state $\rho$ with a probability vector $\mathbf{q}$ on the diagonal may be obtained this way.

%Technically, it would mean to supply a diagonal non-negative trace-one matrix with off-diagonal elements -- in such a way that the new matrix is positive semidefinite and has trace one. 

\begin{figure} [b]
\begin{center}
\includegraphics[width=0.35\textwidth]{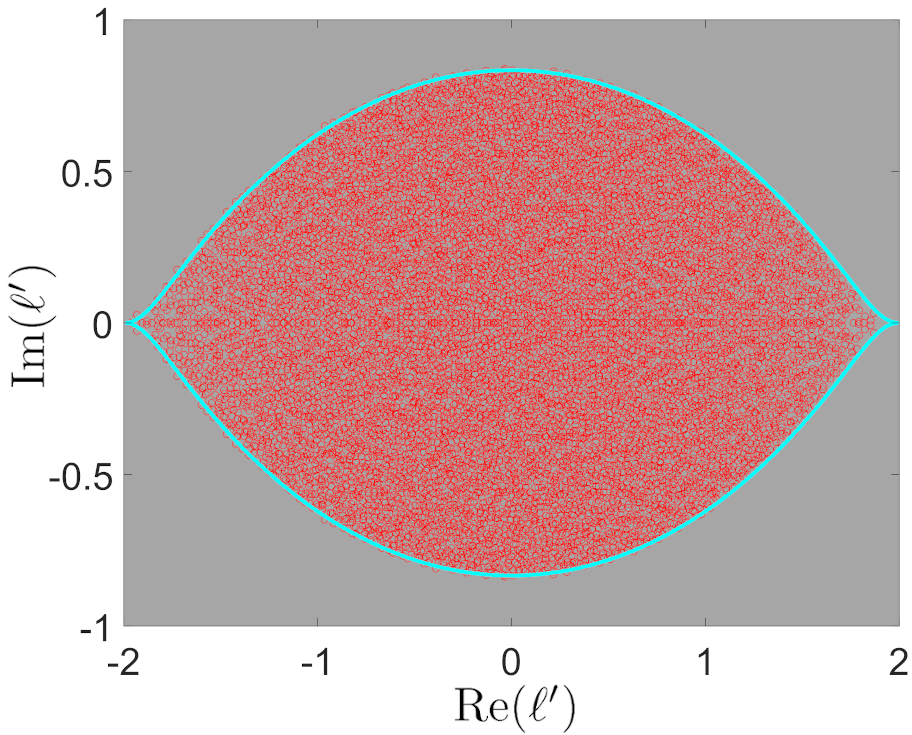}
\end{center}
\caption{Rescaled eigenvalues,	$\ell'= N(\ell +1)$,  of a Lindblad operator 
${\cal L}$ obtained by performing the generic coherification (see text) on a randomly sampled Kolmogorov operator $\cal{K}$ for for $N=100$.~Bright (cyan) contours  corresponds to spectral boundary of random Lindblad generators, Eqs.~(\ref{eq:BorderlineFinalM}-\ref{eq:contour2}).}
\end{figure}

The same procedure may be applied to  maps or, equivalently, to the corresponding Choi matrices. Now it is intuitive how to coherify a classical Kolmogorov generator ${\cal K}_{ij}$ into a quantum Lindbladian.  
%The state we are dealing here with is a Choi state $\mathbf{C}$, Eq.~(\ref{Choi1}).

Starting with the given Kolmogorov generator
$\cal{K}$, see Section \ref{sec:2c},
we can construct a fully decohered Choi matrix $\mathbf{C}^{(0)}$ in the form of a diagonal matrix with elements $\mathbf{C}^{(0)}_{ij,ij} = M_{ij}$. 
%Note that $N$ diagonal elements, $\mathbf{C}^{(0)}_{ii,ii}$, are identically zero. In fact, their values are not important,  since, in the final Lindblad operator ${\cal L}$, the corresponding contributions will be nullified by the second term on the rhs of Eq.~(\ref{L1}).
Next step is a coherefication of the diagonal Choi matrix $\mathbf{C}^{(0)}$. The {\it extreme} coherfication \cite{COH}, i.e., a construction of a pure state $\mathbb{C}^{\rm COH}$, corresponds to $\mathbf{C}^{\rm COH}_{ij,kl} = \sqrt{M_{ij} M_{kl}} e^{i(\phi_{ij} - \phi_{kl})}$, where  phases $\phi_{ij}$ are randomly distributed over the interval $[-\pi,\pi]$. The so obtained completely positive map $\Phi$ is of the form $\Phi(X) = K X K^\dagger$, i.e. it has the Kraus representation with a single Kraus operator. This is a very {\it a}typical map and it leads therefore to an atypical Lindblad operator. The last step is to perform random superdecoherence of $\mathbb{C}^{\rm COH}$ by following Eq.~(\ref{S-DEC}), with a random decoherence matrix $\Delta_{ij,kl}$ defined as follows:

\begin{figure*} [t]
\begin{center}
\includegraphics[width=0.99\textwidth]{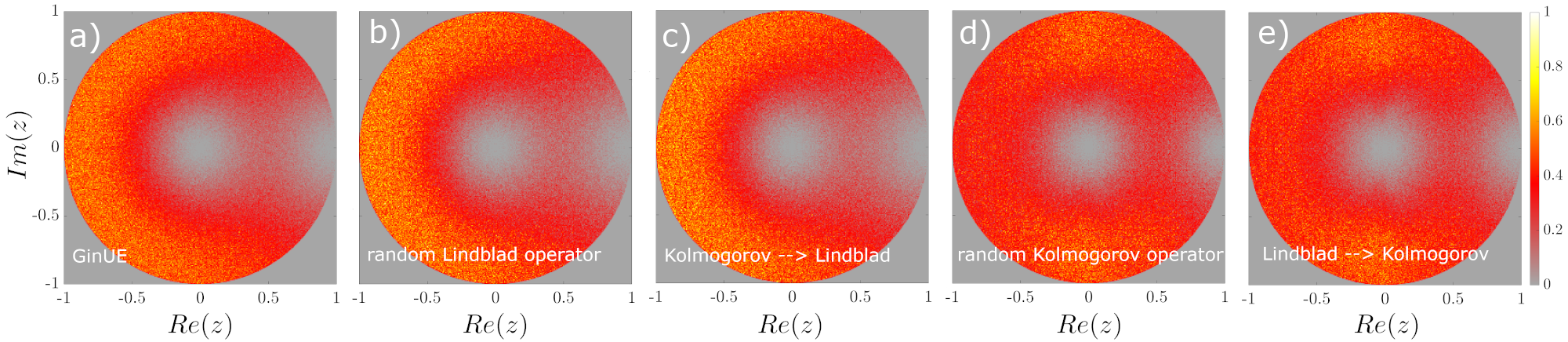}
\end{center}
\caption{Probability density function $P_{\mathrm{CSR}}[\mathrm{Re}(z),\mathrm{Im}(z)]$
of complex level-spacing ratios, Eq.~(\ref{csr}), obtained for 
different ensembles. 
(a) $200$ random $10^4 \times 10^4$ matrices  drawn from the Gaussian Ginibre Unitary Ensemble;  (b) ensemble of random Lindblad operators sampled according to the L-representation, Eq.~(\ref{L1}); (c) ensemble of  random Lindblad operators obtained by coherifying an ensemble of random Kolmogorov operators; (d) ensemble of random Kolmogorov operators; (d) ensemble of Kolmogorov operators obtained by decoherifying an ensemble of random Lindblad operators. The dimension $N=100$ is the same for all the ensembles; however, in order to have equal umber of eigenvalues, Lindblad and Kolmogorov operators were sampled with $200$   and  $20000$ realizations, respectively.}
\end{figure*}

\begin{itemize}
\item[]- sample $N^2 \times N^2$ Wishart matrix $W = GG^\dagger$, where $G$ is drawn from the ensemble of complex $N^2 \times N^2$ Ginibre matrices;
\item[]-construct a diagonal matrix $D_W = diag(1/\sqrt{W_{11}},...,1/\sqrt{W_{N^2N^2}})$ , 
\item[]- define $\Delta := D_W W D_W$.
\end{itemize}
By construction $\Delta > 0$ and $\Delta_{ij,ij}=1$. Hence, 
$C_{ij,kl} := \Delta_{ij,kl} C^{\rm COH}_{ij,kl}$ defines a random partially coherified Choi matrix compatible (on the diagonal) with a classical matrix $M_{ij}$. Note, that $\Delta$ is a Wishart matrix with all diagonal elements equal to 1.

Equivalently, instead of random superdecoherence of $\mathbf{C}^{\rm COH}$ one may sample a Wishart matrix with the given diagonal $W_{ij,ij}=M_{ij}$. 
%However, this is a bad choice because this pure Choi state, after the reshuffling, Eq.~(\ref{resh}), would result in a quantum map of Kraus rank one, $\Phi(\cdot) = K \cdot K^\dagger$, i.e., in an isometry \cite{res2}. This is a very {\it a}typical map and it leads therefore to an atypical Lindblad operator. What is needed is a coherification which allows to sample a typical quantum state with prescribed diagonal entries, i.e., a procedure to sample a Wishart matrix with the given diagonal.
Such procedure is  known in mathematical literature \cite{ignatov,veleva} and it is essentially equivalent to the one based on extremal coherification $\mathbf{C}^0 \to \mathbf{C}^{\rm COH}$ and then random superdecoherence.
Indeed, any $M \times M$ Wishart matrix  $W$ can be represented as a product
$ W = D  V D$,  where $D$ is a diagonal  matrix, $D = diag(\sqrt{\tau_1},... , \sqrt{\tau_M})$ and $V$ is a Hermitian positive definite random matrix with units on the diagonal. Variables $\tau_j$, $j = 1,... ,M$ are mutually independent identically chi-square distributed random numbers, $\tau_j \sim \chi^2$~\cite{ignatov}. Note that this is precisely our case, since, by construction presented in Section~\ref{sec:2c}, the diagonal elements (that are $M_{ij}$) are chi-square distributed. The matrix $D$ can be represented as a correlation matrix obtained by using a multivariate Gaussian distribution~\cite{veleva}.

%Using this result, we define the following {\it generic} coherification:
%\begin{itemize}
%\item[]- sample $N^2 \times N^2$ Wishart matrix $W = %GG^\dagger$, where $G$ is drawn from the ensemble of complex %$N^2 \times N^2$ Ginibre matrices;
%\item[]-construct matrix %$D=diag(\sqrt{W_{11}},...,\sqrt{W_{N^2N^2}})$;
%\item[]-calculate matrix $C' = DWD$;
%\item[]-fill the diagonal of $C'$ with diagonal elements of %$\mathbf{C}$.
%\end{itemize}

%In order to sample random Lindbladians, matrix $C'$ has to be normalized, $\mathrm{Tr} C' = N$, as described in Section \ref{sec:2q}. Choi state $\mathbf{C}'$ corresponds to the case $p=1$. From this state we can descend to any partially coherified state -- by  performing decoherence on $\mathbf{C}'$ with given value $p < 0$.

Figure~11 presents the spectrum of a Lindblad operator obtained by performing the coherefication by using the DVD-decomposition of a randomly sampled Kolmogorov operator.  

\section{Complex spacing ratio statistics of random generators}
\label{sec:5}

So far we considered macroscopic spectral densities of random Lindblad and Kolmogorov operators. Microscopic spectral statistics, which allow to capture correlations between eigenvalues at the scale of typical separation between them,
are also of interest here. More specifically, we ask the question
whether the universality,  observed on the level of (macroscopic) eigenvalues distributions, extends to the microscopic level, where local correlations between the eigenvalues are accounted.

In RMT  and Hamiltonian Quantum Chaos theory, the main tool to quantify correlations between real eigenvalues (energy levels) of Hamiltonian operators is distribution $P(s)$ of spacing between consecutive levels, $s_j = \varepsilon_{j+1} - \varepsilon_{j}$ ~\cite{Mehta}. One of the landmark results  of RMT is the power-law level repulsion $P(s) \propto s^{-\beta}$ in the limit $s \rightarrow 0$, with exponent values specific to the three main Gaussian ensembles. In practice, when dealing with spectra of model Hamiltonians,  in order to eliminate the dependence on fluctuating local energy density and compare the obtained distributions with analytic results, a  complicated  unfolding procedure \cite{Haake1} needs to be performed. It can be avoided if we use \textit{ratios} of consecutive spacing \cite{oganesyan}. Analytic  expressions for spacing ratio (SR) distributions for different RM ensembles were derived \cite{atas} and currently these distributions are popular tools to analyze many-body Hamiltonians; see, e.g., Refs.~\cite{pal,gritsev}.

Recently, the notion of spacing ratios was generalized to the case of non-Hermitian operators~\cite{Prosen-PRX2020}. Namely, for  eigenvalue $\lambda_k$ one has to find, by using the distance on the complex plane as a measure, the  nearest-neighbor, $\lambda^{NN}_k$, and next-to-nearest-neighbor, $\lambda^{NNN}_k$, eigenvalues. The complex spacing ratio (CSR) is then defined as 
\begin{equation}\label{csr}
z_k = \frac{\lambda^{NN}_k - \lambda_k}{\lambda^{NNN}_k - \lambda_k}. 
\end{equation}
CSR values are confined to the unit disc so that
the corresponding probability density function  $P_{\mathrm{CSR}}[\mathrm{Re}(z),\mathrm{Im}(z)]$ has the latter as a support. In Ref.~\cite{Prosen-PRX2020} this distribution was used to categorize different many-body Lindblad operators as ``chaotic" and ``regular" ones. Namely, chaotic Lindblad operators yield CSR distributions similar to the one exhibited by the Gaussian Ginibre Unitary Ensemble (GinUE)~\cite{ginibre}, while regular Lindbladians exhibit CSR distributions characteristic to diagonal matrices with complex Poisson-distributed entries. In the former case the eigenvalues are correlated and this leads to a distinctive horse-shoe pattern with depletion regions near $z=0$ and $z=1$ (see Fig.~12a), while in the latter case the eigenvalues are independent and, in the asymptotic limit $N \rightarrow \infty$, $P_{\mathrm{CSR}}$ is a flat distribution over the unit disc.

We followed the recipe from Ref.~\cite{Prosen-PRX2020} and sampled CSR distributions for different ensembles of operators. As suggested, we only took eigenvalues  from the bulk of the spectral densities and avoided the region near the real axis. Figure~12(b) shows the CSR distribution obtained for random Lindblad operators,
$\lambda_k \equiv \ell_k$ in Eq.~(\ref{csr}), sampled by using L-representation, Eq.~(\ref{L1}), for $N=100$. It has a shape near identical to the one obtained for an ensemble of Gaussian Ginibre Unitary matrices of the size $N^2 =10^4$ ~\cite{comment1}. A similar structure is exhibited [see Fig.~12(d)] by the CSR distribution obtained for an ensemble of  Kolmogorov operators, $\lambda_k \equiv \chi_k$ in Eq.~(\ref{csr}). However, in this case the deviation from the CSR distribution presented in Fig.~12(a) is more pronounced; we attribute this to finite-size effects.

It is not a surprise that random  Lindblad 
and Kolmogorov generators exhibit CSR distributions similar to the the one obtained for the GinUE ensemble. What is  reamarkable is that both procedures, superdecoherence (Section V) and supercoherification (Section VI), preserve this property. Figure 12(c) shows $P_{\mathrm{PCS}}$ sampled with an ensemble of Lindbladians obtained from an ensemble of random Kolmogorov operators by performing the $DVD$ procedure (see Section VI). The so obtained distribution is identical (within the sampling error) to the one obtained with the straightforward sampling, Fig.~12(b). A similar result is observed with  Kolmogorov operators obtained by decoherefying ensemble of random Lindbladians. Note that in this case even the finite-size effects are reproduced  (see, e.g.,  the shape of the depletion region near $z=0$). 

\section{Conclusions}\label{sec:6}

In this work we analyzed the spectra of random Lindblad generators
and their classical analogues,  Kolmogorov generators. This work extends earlier  results  \cite{denysov2019}
on the support of the
spectrum of random Lindblad operators by evaluating the probability density 
on the complex plane, which is one of our main results. The second main results is the analysis of the  quantum--to--classical 
transition at the level of  generators of continuous time Markovian dynamics, induced by superdecoherence. The strength of superdecoherence is characterized by a single parameter $p$, interpolating between $0$ (complete decoherence) and $1$ (zero decoherence).

In particular,  we show that
the quantum-to-classical transition is size-dependent. The transition is sharp and takes place at $p_{tr}$, which scales with $N$ as  $p_{tr} \propto \frac{1}{\sqrt{N}}$. In other words, as the system size $N$ increases, the transition to the classical regime (at least in terms of the spectral density) happens closer and closer to  the point $p=0$ (complete decoherence). What happens to the eigen-elements of a random Lindbladian during this transition, is an interesting question. Is their transformation going faster or slower? If the latter, then a {\em typical} Lindbladian, governing Markovian evolution in a very large Hilbert space, is able to withstand very strong decoherence and remain a typical Lindblad operator without losing its quantum features.

The results presented in our work
is hardly applicable to describe spectral
properties of a particular physical system. However, situation changes
if one considers an ensemble of
quantum (or classical) systems, averaged over a suitably chosen set of parameters.
In such a case the distributions derived in this work  provides a fair approximation of average spectral properties of such an ensemble of physical systems -- 
if  the corresponding classical dynamics is strongly chaotic and the coupling of the system with an environment is strong  enough.

We did not consider random Lindblad operators with a non-zero unitary part. Note, however, that  term $\mathcal{L}_H(\rho) = -i[H,\rho]$ gives rise to the following Choi matrix,
\begin{equation}
    \mathbb{C}^H_{mn,kl} = -i(H_{nm}\delta_{kl} - H_{kl}\delta_{mn}) , 
\end{equation}
and hence the diagonal elements
\begin{equation}
    \mathbb{C}^H_{kl,kl} = -i(H_{lk} - H_{kl})\delta_{kl} = 0,
\end{equation}
do not contribute to the Kolmogorov generator. Interestingly, the super-decoherence of off-diagonal elements $\mathbb{C}^H_{mn,kl} \to p \mathbb{C}^H_{mn,kl}$ corresponds to a simple scaling of the Hamiltonian, $H \to pH$, and therefore this operator vanishes in the classical limit, $p=0$. The effect of superdecoherence is different if we first find the unitary evolution, $e^{t \mathcal{L}_H}\rho = U(t) \rho U^\dagger(t)$ with $U(t) = e^{-iHt}$, and then allow for superdecoherence. One finds the diagonal elements of the Choi matrix, 
\begin{equation}
      \mathbb{C}^U_{kl,kl} = |U_{kl}(t)|^2 ,
\end{equation}
which defines a doubly stochastic matrix. This reasoning  clearly shows
that the two operation, $\mathcal{L} \to e^{t \mathcal{L}}$ and superdecoherence, do not commute. 

%While it is more or less intuitive that, in the limit of complete decoherence, an ensemble of such quantum generators would also end up as an ensemble of random Kolmogorov generators, the classical-to-quantum transition can be essentially different from what we observed with purely dissipative Lindbladians.

Finally,  complex spacing ratio statistics are also warrant a further study. It is an interesting question whether superdecoherence (coherification) can modify the PCS distribution
in a such a way that the corresponding generator changes its type, e.g., from "chaotic" to "regular". 

%Knowing the support of the spectrum of the %Lindblad or  the Kolmogorov
%generators, we can estimate the size of the %spectral gap of the system.
%This key quantity allows one to characterize %the expected 
%rate of convergence to the equilibrium, 
%which changes during the quantum-to-classical %transition.

\section{Acknowledgments}\label{acknowledgment}

It is a pleasure to thank
Boris Khoruzhenko,
{\L}ukasz Pawela and Zbigniew Pucha{\l}a
for several discussions and helpful remarks. 
%on the process of superdecoherence.
%    The authors acknowledge support of the 
The numerical
experiments and simulations were supported by the Russian Science Foundation Grant No. 19-72-20086 (I.Y., T. L., and S. D.) and
were performed on the supercomputer “Lomonosov-2” of the Moscow State University.
Financial support by Narodowe Centrum Nauki under the grants number 2018/30/A/ST2/00837  (DC),
DEC-2015/18/A/ST2/00274  (K{\.Z}),
by the Foundation for Polish Science under the Team-Net 
project no. POIR.04.04.00-00-17C1/18-00 is acknowledged.
T. L. acknowledges support by the Basis Foundation (Grant No. 18-1-3-66-1). S. D. acknowledges the support by NordSTAR - Nordic Center for Sustainable and Trustworthy AI
Research (OsloMet Project Code 202237-100).

%\section{Appendices}

\renewcommand\thefigure{\thesection.\arabic{figure}}  

\appendix
\section{Evaluation of the spectral densities}
\setcounter{figure}{0}   
%\subsection{Appendix A: Evaluation of the %universal spectral distribution}
%\label{sec:AppA}

We start with a brief review on the quaternionic extension of free probability to non-Hermitian random matrices, developed in Refs.~\cite{S2,S3,S4,S5,S6}
(see also Ref.~\cite{S7}), focusing mostly on the aspects relevant to our analysis. For a more detailed introduction and  explicit calculations we refer to Ref.~\cite{S8}.

The main  object of our interest is the spectral density $\rho(z,\zb)=\left<\frac{1}{N}\sum_{i=1}^{N}\delta^{(2)}(z-\lambda_i)\right>$ on the complex plane. Here $\delta^{(2)}(z)=\delta(\textup{Re} z)\delta(\textup{Im} z)$. The density is obtained via the Poisson law $\rho(z,\zb)=\lim_{|w|\to 0}\frac{1}{\pi}\partial_{z\zb}\Phi(z,\zb,w,\wb)$, where $\Phi$ is the (regularized) electrostatic potential in two dimensions \cite{S9},
%~\cite{Sompolinsky}
%\begin{eqnarray}
%    \Phi(z,\zb,w,\wb)= ~~~~~~~~~~~~~~~\nonumber \\
%    \<\frac{1}{N}\ln \det\left[(z-X)(\zb-X^{\dagger})+|w|^2\right]\>.
%\end{eqnarray}
\begin{multline}
    \Phi(z,\zb,w,\wb)=  \\
    \<\frac{1}{N}\ln \det\left[(z-X)(\zb-X^{\dagger})+|w|^2\right]\>.
\end{multline}
To facilitate the calculations in the large $N$ limit, we consider the generalized Green's function, which is a $2\times 2$ matrix
\begin{equation} \nonumber
    \cG(Q)=\left<\frac{1}{N}\btr \left(Q\otimes \oper-\cX\right)^{-1}\right>=\left(\begin{array}{cc}
    \cG_{11}     & \cG_{12} \\
    \cG_{21}     & \cG_{22}
    \end{array}\right),
\end{equation}
with
\begin{equation}
Q=\left(\begin{array}{cc}
z       & i\wb \\
    iw & \zb
\end{array} \right),\quad
\cX=\left(\begin{array}{cc}
    X & 0 \\
    0  & X^{\dagger}
\end{array}\right),\label{eq:QuatDef}
\end{equation}
where we also introduced a block trace (partial trace) operation
\begin{equation}
\btr\left(\begin{array}{cc}
A     & B \\
C     & D
\end{array}\right)=\left(\begin{array}{cc}
    \textup{Tr} A & \textup{Tr} B \\
    \textup{Tr} C & \textup{Tr} D
\end{array}\right).
\end{equation}
Note that $Q$ is the matrix representation of a quaternion, thus we refer to this approach as quaternionic free probability. The upper-left element of $\cG$ yields spectral density via $\rho(z,\zb)=\lim_{|w|\to 0}\frac{1}{\pi}\partial_{\zb}\cG_{11}$, while the product of off-diagonal elements yields the correlation function
capturing non-orthogonality of eigenvectors~\cite{S12},
associated with left ($\langle L_i |$) and right ($| R_i \rangle$) eigenvectors $O(z,\zb)=-\frac{1}{\pi}\lim_{|w|\to 0}\cG_{12}\cG_{21}$, where~\cite{S10,S11}
%\begin{eqnarray}
%O(z,\zb)= ~~~~~~~~~~~~~~~~~~~~~~~\nonumber \\
%\lim_{N\to\infty} \<\frac{1}{N^2}\sum_{i=1}^{N}\langle L_i|L_i\rangle\langle R_i|R_i \rangle \delta^{(2)}(z-\lambda_i)\>.
%\label{eq:O1}
%\end{eqnarray}
\begin{multline}
O(z,\zb)=  \\
\lim_{N\to\infty} \<\frac{1}{N^2}\sum_{i=1}^{N}\langle L_i|L_i\rangle\langle R_i|R_i \rangle \delta^{(2)}(z-\lambda_i)\>.
\label{eq:O1}
\end{multline}
An important fact  is that the boundary of the spectrum can be determined from the condition $\cG_{12}\cG_{21}=0$.

Knowing the Green's function, we can define also the Blue's function as its functional inverse,
\begin{equation}
\cB\big(\cG(Q)\big)=Q,\quad \cG\big(\cB(Q)\big)=Q. \label{eq:GBrel}
\end{equation}
Then, the quaternionic $R$-transform is defined as $\cR(Q)=\cB(Q)-Q^{-1}$, where the inverse is understood in the sense of $2\times 2$ matrix inversion. When two non-Hermitian matrices $A$ and $B$ are free, then the $R$-transform of their sum is a sum of corresponding $R$-transforms
\begin{equation}
\cR_{A+B}(Q)=\cR_A(Q)+\cR_B(Q). \label{eq:AdditionLaw}
\end{equation}
In that sense, it generalizes the logarithm of the Fourier transform from classical probability to the noncommutative case.

We now consider a problem of finding the spectrum of a matrix $A+B$, where $A$ is a Ginibre matrix and $B$ can be arbitrary. Starting with Eq. (\ref{eq:AdditionLaw}), we add $Q^{-1}$ to both sides, obtaining
\begin{equation}
\cB_{A+B}(Q)=\cR_A(Q)+\cB_B(Q).
\end{equation}
Then we make a substitution $Q\to\cG_{A+B}(Q)$ and use the relation between Green's and Blue's function, Eq.~(\ref{eq:GBrel}), obtaining
\begin{equation}
Q-\cR_A\big(\cG_{A+B}(Q)\big)=\cB_B\big(\cG_{A+B}(Q)\big).
\end{equation}
Next we evaluate the Green's function of $B$ on both sides of the equation and
by using Eq.~(\ref{eq:GBrel}), obtain
\begin{equation}
\cG_B\Big(Q-\cR_A\big(\cG_{A+B}(Q)\big)\Big)=\cG_{A+B}(Q), \label{eq:PasturEq}
\end{equation}
which is the non-Hermitian Pastur equation.
In our case $A$ is Ginibre, the $R$-transform of which reads
\begin{equation}
\cR_{A}(\cG_{A+B})=\left(\begin{array}{cc}
    0 & \cG_{12} \\
    \cG_{21} & 0
\end{array}\right),
\end{equation}
thus \eqref{eq:PasturEq} simplifies to
\begin{equation}
\cG_{B}\left[\left(\begin{array}{cc}
z     & -\cG_{12} \\
 -\cG_{21}     & \zb
\end{array}\right)\right]=\left(\begin{array}{cc}
\cG_{11}     & \cG_{12} \\
\cG_{21}     & \cG_{22}
\end{array}\right), \label{eq:PasturSimp}
\end{equation}
where we suppressed the index `$A+B$' when writing components of $\cG_{A+B}$.
We also used the fact that all important quantities are calculated in the $|w|\to 0$ limit and took this limit at the level of this algebraic equation.

%\subsection{Embedding of hermitian matrices}
Equation (\ref{eq:PasturSimp}) holds for general (not necessarily random) matrix $B$. In our case of Lindblad and Kolmogorov generators, $B$ is Hermitian, which simplifies the calculation of its quaternionic Green's function. It  reads \cite{S6}
%~\cite{JaroszNowak}
\begin{equation}
    \cG(Q)=\gamma(q,\bar{q})\oper_2-\gamma'(q,\bar{q})Q^{\dagger}, \label{eq:Gherm}
\end{equation}
with
\begin{eqnarray}
\gamma(q,\bar{q})=\frac{q G(q)-\bar{q}G(\bar{q})}{q-\bar{q}}, \nonumber \\ %\\qquad
\gamma'(q,\bar{q})=\frac{G(q)-G(\bar{q})}{q-\bar{q}}, \label{eq:Gammas}
\end{eqnarray}
where $q,\bar{q}$ are the eigenvalues of the $2\times 2$ quaternion matrix (\ref{eq:QuatDef}) and $G(z)$ is the Stieltjes transform of the spectral density of $B$
\begin{equation}
G(z)=\int_{-\infty}^{+\infty}\frac{\rho_B(x)dx}{z-x}. \label{eq:Greens}
\end{equation}

\begin{figure}[t]
\begin{center}
\includegraphics[width=0.4\textwidth]{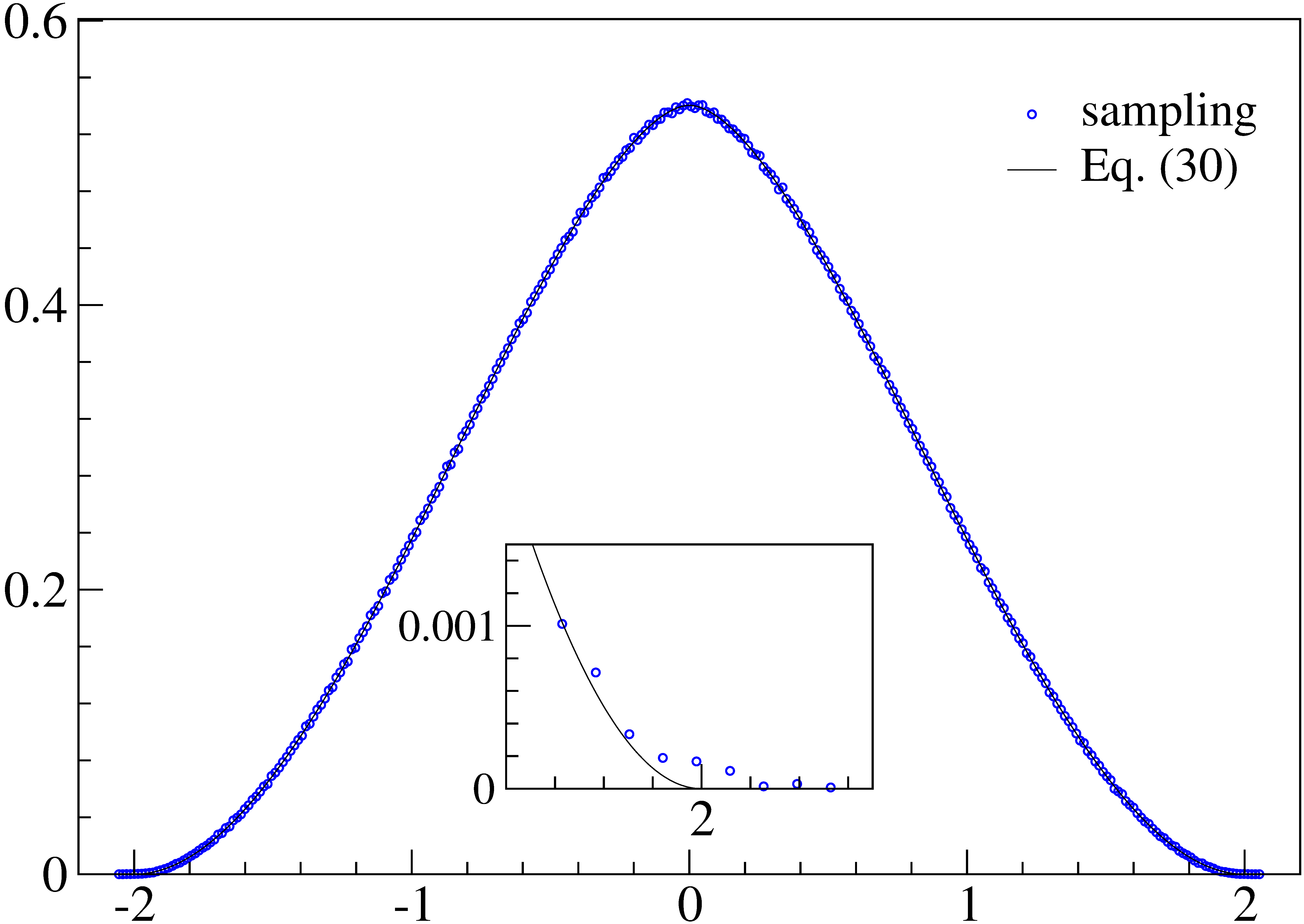}
\end{center}
\caption{Spectral density of the
matrix $B = \oper \otimes C + C \otimes \oper$,
where $C$ is a $N \times N$ GOE matrix
and $\oper$ is a $N \times N$ identity matrix. Analytical asymptotic result, Eq.~(\ref{eq:rhoB}) (black line) is in a good agreement with the result of a sampling for $N=100$ (blue circles). Total number of samples is $10^3$.
\label{Fig:SpecB}
}
\end{figure}

%\subsection{Border of the spectrum} \label{sec:Border}
We are now ready to solve Eq. (\ref{eq:PasturSimp}). The quaternion matrix of our interest is now
\begin{equation}
Q=\left(\begin{array}{cc}
z     & -\cG_{12} \\
-\cG_{21}     & \zb
\end{array}\right). \label{eq:AuxMat}
\end{equation}
We denote its complex conjugate eigenvalues as $q$ and $\bar{q}$.

Focusing on the upper-right component of the matrix equation \eqref{eq:PasturSimp} and using $\bar{\cG}_{21}=-\cG_{12}$, which follows from the definition of the quaternion, Eq.~(\ref{eq:QuatDef}), we obtain
\begin{equation}
-\frac{G(q)-G(\bar{q})}{q-\bar{q}}\cG_{12}=\cG_{12}.\label{eq:Bord1}
\end{equation}
%Now $q,\bar{q}$ are the eigenvalues of the matrix
%\begin{equation}
%\left(\begin{array}{cc}
%z     & -\cG_{12} \\
%-\cG_{21}     & \zb
%\end{array}\right). \label{eq:AuxMat}
%\end{equation}
There is one trivial solution, $\cG_{12}=0$, which corresponds to vanishing of the eigenvector correlation function, Eq.~\eqref{eq:O1}. This solution is valid outside of the spectrum, simply because there are no eigenvalues contributing to Eq.~\eqref{eq:O1}. Inside the spectrum $\cG\neq 0$, thus one has 
\begin{equation}
G(q)-G(\bar{q}) = \bar{q}-q. \label{eq:Mainq}
\end{equation}
This equation for the complex variable $q$ equates only imaginary parts, therefore solutions form a one-dimensional set on the complex plane. Moreover, it is invariant under the interchange $q\leftrightarrow \bar{q}$, thus the solutions to Eq.~\eqref{eq:Mainq} come in complex conjugate pairs.
To relate them with the actual position on the complex plane, let us recall that $q$ and $\bar{q}$ are the eigenvalues of the matrix in Eq.~\eqref{eq:Mainq}, thus $q + \bar{q} = \textup{Tr} Q = z+\bar{z}$. Using the second matrix invariant, we obtain $|q|^2 = \det Q = |z|^2-\cG_{12}\cG_{21}$. This immediately leads us to the formula for the eigenvector correlation function $O(z,\bar{z}) = \frac{1}{\pi}(|z|^2-|q|^2)$. This correlation function vanishes at the boundary of the spectrum, so we immediately conclude that $q=z$ at the boundary. Therefore, the boundary of the spectrum can be derived from the condition 
\begin{eqnarray}
G(z) - G(\zb) + z -\zb = 0. \label{eq:BorderlineMain}
\end{eqnarray}

To find the spectral density, we focus on the upper-left component of Eq. \eqref{eq:PasturSimp}. Using Eq. \eqref{eq:Gherm} and Eq. \eqref{eq:Mainq}, we find $\cG_{11} = \zb + \gamma(q,\bar{q})$. The spectral density is then given by $\rho(z,\zb) = \frac{1}{\pi} + \frac{1}{\pi}\partial_{\zb} \gamma(q,\bar{q})$. It only remains to calculate the derivative of $\gamma$, which is given by Eq. \eqref{eq:Gammas}. As an intermediate step, derivatives of $q$ and $\bar{q}$ can be found by differentiating the relation $q+\bar{q} =  z+\zb$ and Eq. \eqref{eq:Mainq}. The spectral density finally reads
\begin{widetext}
\begin{eqnarray}
\rho(z,\zb) = \frac{1}{\pi} + 
%\frac{1}{\pi}
\frac{(G(q)+qG'(q))(1+G'(\bar{q})) - (G(\bar{q})+\bar{q} G'(\bar{q}))(1+G'(q))}{\pi (q-\bar{q})(2+G'(q)+G'(\bar{q}))} + 
%\nonumber
%\\
%\frac{1}{\pi}
\frac{(qG(q)-\bar{q}G(\bar{q}))(G'(q)-G'(\bar{q}))}{\pi (q-\bar{q})^2(2+G'(q)+G'(\bar{q}))}.
\label{eq:DensityComplex}
\end{eqnarray}
\end{widetext}
We recall that $q$ and $\bar{q}$ are the solutions of Eq. \eqref{eq:Mainq} and are related with the position on the complex plane by $\textup{Re}\, q = \textup{Re}\, z$. There is no relation between $q$ and the imaginary part of $z$, thus the spectral density is independent of $\textup{Im}\, z$ inside the region bounded by the solutions of Eq.~\eqref{eq:BorderlineMain}.
These considerations apply to any Hermitian matrix $B$, but in this work we focus on two cases corresponding to Lindblad and Kolmogorov generators for which we explicitly calculate Green's functions.

In the case of purely dissipative Lindbladian $B=1\otimes C+C \otimes 1$, where $C$ is a GOE matrix, the spectrum of which is the Wigner semicircle, $\rho_C(x)=\frac{2}{\pi}\sqrt{1-x^2}$ (see Eq.\eqref{eq:SMmodel}).
Note that each eigenvalue of $B$ is of the form $\lambda=\mu_a+\mu_b$, where $\mu_{a,b}$ are the eigenvalues of $C$, so the spectrum of $B$ is the (classical) convolution of two Wigner semicircles, which can be calculated using standard tools from probability. The Fourier transform of the Wigner semicircle reads $\tilde{\rho}_C(k)=\frac{2}{k}J_{1}(k)$, where $J_1$ is the Bessel function of the first kind. Therefore, the Fourier transform of $B$ reads $\tilde{\rho}_B(k)=\frac{4}{k^2}J_1^2(k)$.
The Fourier transform can be inverted, yielding the spectral density
\begin{eqnarray}
\rho_B(x)=\frac{|x|}{\pi}\MeijerG{0}{2}{2}{2}{1,2}{-\frac{1}{2},\frac{1}{2}}{\frac{x^2}{4}}\chi_{-2\leq x\leq 2},
\label{eq:rhoB}
\end{eqnarray}
where $\MeijerG{m}{n}{p}{q}{a}{b}{x}$ is the Meijer G-function and $\chi_{A}=1$ when $A$ is true and 0 otherwise. The formula above is juxtaposed with the numerical simulation and plotted in Fig.~\ref{Fig:SpecB}.

%\begin{figure}[b]
%\begin{center}
%\includegraphics[width=0.49\textwidth]{fig6.png%}
%\end{center}
%\caption{Overlaps between the spectral %distributions used to calculate Jaccard index.
%\label{fig:12}
%}
%\end{figure}

To evaluate the Green's function, defined by Eq.~\eqref{eq:Greens}, we use the following representation $(z-x)^{-1}=\mp i\int_0^{\infty}e^{\pm ik(z-x)}dk$, which allows us to calculate the Stieltjes transform directly from the Fourier transform via $G(z)=\mp i\int_{0}^{\infty}e^{\pm ikz}\tilde{\rho}_B(\mp k)dk$, where we take the upper signs for $\textup{Im}\, z>0$ and lower for $\textup{Im}\, z<0$. The final result reads
\begin{equation}
G(z)= 2z-\frac{2z}{3\pi}\left[(4+z^2)E\left(\frac{4}{z^2}\right)+(4-z^2)K\left(\frac{4}{z^2}\right)\right] \nonumber
\label{eq:contour2S}
\end{equation}
where $K(z)$ and $E(z)$ are the complete elliptic integrals of the first and second kind, respectively.

%Then we substitute the result into equation~(\ref{eq:BorderlineFinal}) and obtain an implicit equation which is then can be solved numerically. The corresponding solution is shown on Fig.~1 (bright contour).

In the case of Kolmogorov generators, the matrix $B$ is diagonal with Gaussian elements, thus the evaluation of the Green's function is straightforward
%\begin{widetext}
%\begin{equation}
%G(z) = \frac{1}{\sqrt{2\pi}}\int_{-\infty}^{\infty} \frac{e^{-x^2/2}}{z-x}dx =
%\sqrt{\frac{\pi}{2}}e^{-z^2/2}\left(\textup{Erfi}\left(\frac{z}{\sqrt{2}}\right)-i \textup{sgn} (\textup{Im} z)\right), \nonumber
%\end{equation}
%\end{widetext}

\begin{multline}
G(z) = \frac{1}{\sqrt{2\pi}}\int_{-\infty}^{\infty} \frac{e^{-x^2/2}}{z-x}dx = 
\\
\sqrt{\frac{\pi}{2}}e^{-z^2/2}\left(\textup{Erfi}\left(\frac{z}{\sqrt{2}}\right)-i \textup{sgn} (\textup{Im} z)\right), \nonumber
\end{multline}
where $\textup{Erfi}(z)=-i\textup{Erf}(iz)$ and $\textup{Erf}(z)=\frac{2}{\sqrt{\pi}}\int_0^z e^{-t^2}dt$ is the error function.

\section{Intersection over Union and  distance between two spectral borders}
\setcounter{figure}{0}   

The Jaccard index \cite{jacck1,jacck2} quantifies similarity between finite  sets, and is defined as the size of the intersection divided by the size of the union of the sets. In the case of two contours, $A$ and $B$ (that are spectral borders in our case), size is given by the corresponding area, and we have what is also called 'Intersection over Union' (that is a short-hand version of 'area of the intersection/overlap over the area of union'),
\begin{equation}
\mathrm{IoU}(A,B) = \frac{\mathrm{area}(A \cap B)}{\mathrm{area}(A \cup B)}.
\end{equation}
If the contours are identical, we have $\mathrm{IoU}(A,B) = 1$. In case they are so different that the overlap between them is zero, we have $J(A,B) = 0$. Therefore, the distance is $d(A,B)=1-\mathrm{IoU}(A,B)$.

Figure B.1 illustrates the idea. The spectral boundaries for sampled distributions were obtained by using MATLAB function $convhull$ which constructs convex hull for the given planar set of points. Before that, the sampled spectral distributions were scaled with  classical and quantum scalings, respectively (see Section V of the main text).

\begin{figure}[b]
\begin{center}
\includegraphics[width=0.49\textwidth]{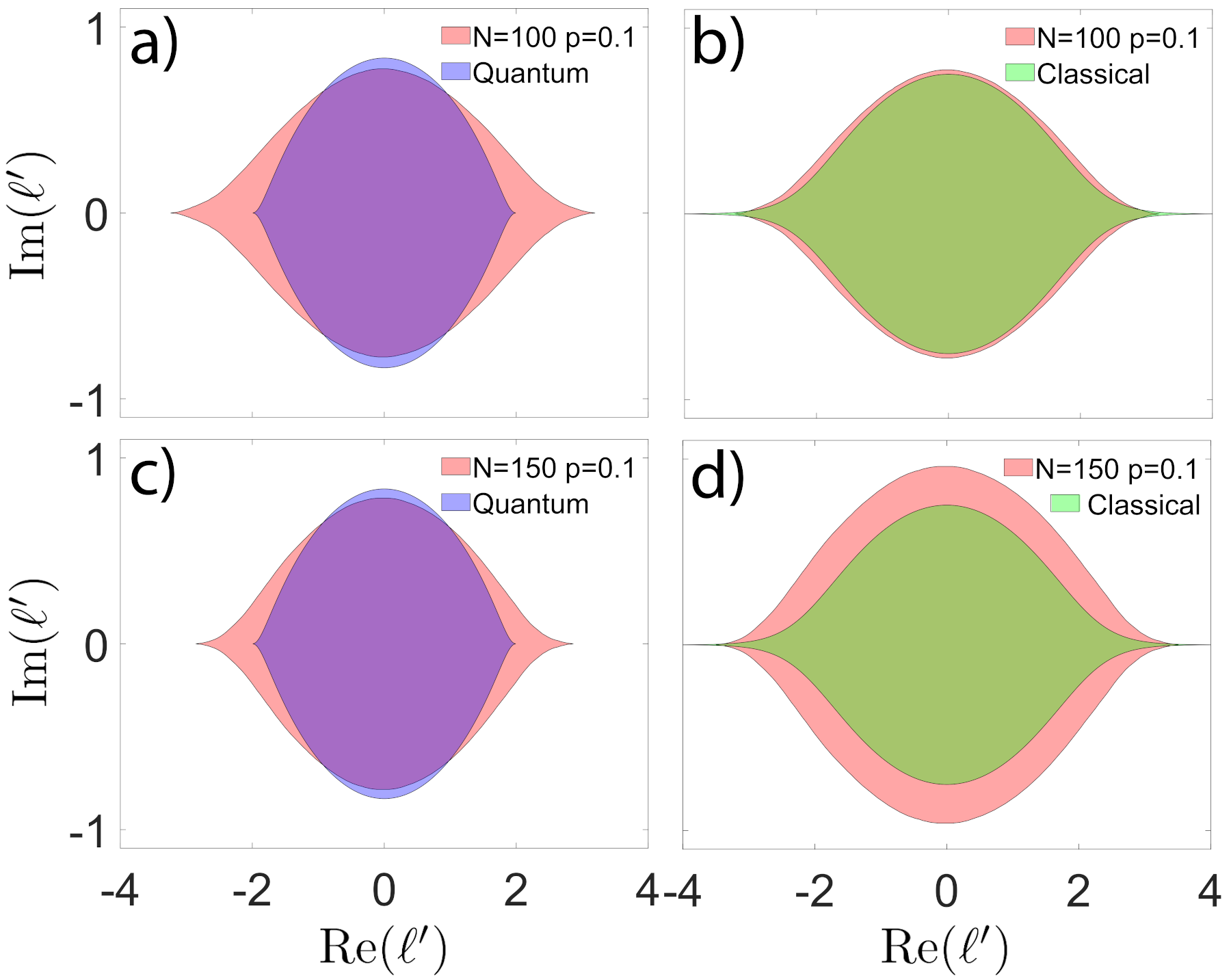}
\end{center}
\caption{Comparison of the spectral borders with classical and quantum contours for $p=0.1$ and two different values of $N$, $100$ and $150$. The sampled eigenvalues are scaled, $\ell'= w(N)(\ell +1)$, with $w(N) = N/p$ (quantum) and $w(N) = N^{\frac{3}{2}}$ (classical).
\label{Fig:B1}
}
\end{figure}

%\newpage


\begin{thebibliography}{99}
\bibitem{sample} For instance, on the supercomputer `Lomonosov 2' \cite{lomonosov}, we were able to reach $N=350$.
\end{thebibliography}
\end{document}